\documentclass[twocolumn]{aastex6}

\usepackage{natbib}
\usepackage{amsmath}
\usepackage{multirow}
\usepackage{graphicx}
\usepackage{color}
\usepackage{threeparttable}
\usepackage{enumitem}
\usepackage{mathtools}
\usepackage{lipsum}

\def\gx339{GX~339--4}

\def\rxte{{\it RXTE}}
\def\xmm{{\it XMM-Newton}}
\def\suzaku{{\it Suzaku}}
\def\swift{{\it Swift}}
\def\nustar{{\it NuSTAR}}
\def\nicer{{\it NICER}}

\def\xspec{{\tt XSPEC}}

\begin{document}
\title{Relativistic reflection and reverberation in \gx339\ with
\textit{NICER} and \textit{N\MakeLowercase{u}STAR}}
\author{Jingyi~Wang\altaffilmark{1},
Erin~Kara\altaffilmark{1},
James~F.~Steiner\altaffilmark{1,2},
Javier~A.~Garc{\'\i}a \altaffilmark{3,4}, 
Jeroen~Homan \altaffilmark{5,6},
Joseph~Neilsen\altaffilmark{7},
Gr\'egoire~Marcel\altaffilmark{7},
Renee~M.~Ludlam\altaffilmark{3}, 
Francesco~Tombesi\altaffilmark{8,9,10,11}, 
Edward~M.~Cackett\altaffilmark{12}, 
Ron~A.~Remillard\altaffilmark{1}.
}
\affil{
\altaffilmark{1}MIT Kavli Institute for Astrophysics and Space
Research, MIT, 70 Vassar Street, Cambridge, MA 02139, USA\\
\altaffilmark{2}Harvard-Smithsonian Center for Astrophysics, 60 Garden St., Cambridge, MA 02138, USA \\
\altaffilmark{3}Cahill Center for Astronomy and Astrophysics, California
Institute of Technology, Pasadena, CA 91125, USA\\ 
\altaffilmark{4}Remeis Observatory \& ECAP, Universit\"{a}t
Erlangen-N\"{u}rnberg, 96049 Bamberg, Germany\\ 
\altaffilmark{5}Eureka Scientific, Inc., 2452 Delmer Street, Oakland, CA 94602, USA\\
\altaffilmark{6}SRON, Netherlands Institute for Space Research, Sorbonnelaan 2, 3584 CA Utrecht, The Netherlands\\
\altaffilmark{7}Department of Physics, Villanova University, 800 Lancaster Avenue, Villanova, PA 19085, USA\\
\altaffilmark{8}Department of Astronomy, University of Maryland, College Park, MD, 20742, USA \\
\altaffilmark{9}X-ray Astrophysics Laboratory, NASA/Goddard Space Flight Center, Greenbelt, MD, 20771, USA \\
\altaffilmark{10}Department of Physics, University of Rome ``Tor Vergata", Via della Ricerca Scientifica 1, 00133, Rome, Italy \\
\altaffilmark{11}INAF - Astronomical Observatory of Rome, via Frascati 33, 00044, Monte Porzio Catone (Rome), Italy\\
\altaffilmark{12}Department of Physics \& Astronomy, Wayne State University, 666 W. Hancock St, Detroit, MI 48201, USA\\
}

\begin{abstract}
We analyze seven \nicer\ and \nustar\ epochs of the black
hole X-ray binary \gx339\ in the hard state during its two most recent hard-only outbursts in 2017 and 2019. These observations cover the $1-100$~keV unabsorbed luminosities between $0.3\%$ and $2.1\%$ of the Eddington limit. With \nicer's negligible pile-up, high count rate and unprecedented time resolution, we perform a spectral-timing analysis and spectral modeling using relativistic and distant reflection models. Our spectral fitting shows that as the inner disk radius moves inwards, the thermal disk emission increases in flux and temperature, the disk becomes more highly ionized and the reflection fraction increases. This coincides with the inner disk increasing its radiative efficiency around $\sim$1\% Eddington. We see a hint of hysteresis effect at $\sim0.3\%$ of Eddington: the inner radius is significantly truncated during the rise ($>49R_{g}$), while only a mild truncation ($\sim5R_g$) is found during the decay. At higher frequencies ($2-7$~Hz) in the highest luminosity epoch, a soft lag is present, whose energy dependence reveals a thermal reverberation lag, with an amplitude similar to previous findings for this source. We also discuss the plausibility of the hysteresis effect and the debate of the disk truncation problem in the hard state. 
\end{abstract}





\section{Introduction} \label{introduction}

Black hole astrophysics can be regarded as a fundamental tool in providing information about the accretion and ejection physics in the strongest gravity regime in the Universe. The standard picture for an accreting black hole system involves an accretion disk that emits as a multi-temperature blackbody ($\sim0.1-2$~keV), and a hot (hundreds of keV) plasma called an X-ray ``corona" whose nature is still not clear (see \citealt{done2007modelling} for a review). Inverse Compton scattering of the thermal photons from the accretion disk off free electrons in the corona generates a Comptonization component, usually modeled by a power-law spectrum with high energy cut-off. A fraction of the Comptonized photons may shine back onto the disk. The interaction of these photons with the material in the accretion disk, including Compton scattering, photoelectric absorption followed by fluorescent line emission or Auger de-excitation, produces a reflection spectrum \citep{garcia2014improved, bambi2017black}. If the reflection happens very close to the black hole, then the local spectrum is expected to be smeared by relativistic effects \citep{fabian1989x}. For this reason, X-ray reflection spectroscopy provides a powerful diagnostic tool for investigating the dynamics and geometry of the accretion disk. The modeling of reflection features is being used to measure the spin, inclination angle, and ionization in a variety of black hole systems (see \citealt{reynolds2013spin} for a review). 

The brightest outburst of black hole binaries (BHBs) can be described by a hysteresis pattern in the hardness-intensity diagram (HID, e.g. \citealt{fender2004towards}, and see the lower panel in Fig.~\ref{fig:lc} (\textit{lower})). The majority of BHBs spend most of the time in the quiescent state when the accretion rate onto the black hole is low and the X-ray emission is weak, often undetected. The X-ray emission is dominated by the Comptonization component in the hard state, while the luminosity increases until it starts its transition to the soft state where the thermal disk component dominates. The luminosity gradually drops and the source makes the transition back to the hard state, and then to quiescence \citep{remillard2006x}. Sometimes, the outbursts are hard-only when the BHB stays in the hard state and the transition to the soft state does not take place \citep{tetarenko2016watchdog}. 

It is of central importance to determine the evolution of physical properties in the accretion disk and the corona, because it could provide us with insights into the accretion process, the nature of the corona and jet, and the mechanisms governing state transitions. 

\gx339\ is a low-mass X-ray binary (LMXRB) that goes into outburst cycles typically every 2-3 years \citep{tetarenko2016watchdog}. A near-infrared study in \citet{heida2017potential}
has constrained the mass of the black hole to $2.3M_\odot<M_{\rm BH}<9.5M_\odot$; the distance to \gx339\ is difficult to be accurately measured, and only a lower limit of $\sim5$~kpc can be derived. In this work, to estimate the luminosity in units of Eddington limit, the distance is assumed to be 8~kpc, and the mass is 10~$M_\odot$ to make comparison with previous results in literature more conveniently. 

In \citet{wang2018evolution}, we analyzed the X-ray spectra of this source during the 2013 and 2015 outbursts with \textit{Nuclear Spectroscopic Telescope Array} (\nustar, \citealt{harrison2013nuclear}) and \textit{the Neil Geherls Swift Observatory} (\swift). In this paper, we present new analysis of the 2017 and 2019 outbursts with \nustar\ and \textit{Neutron Star Interior Composition Interior Explorer (NICER}, \citealt{gendreau2016neutron}). These two recent outbursts are hard-only outbursts because \gx339\ did not make the transition from the hard state to the soft state as it does during its brightest outbursts. \citet{garcia20192017} recently analyzed the 2017 outburst using \nustar\ and complementary \swift\ data. Their best-performing spectral models suggest that an approximation of the corona by two lamppost illuminators offered a better description of the reflection and continuum data than the usually-adopted lamppost plus distant neutral reflection. 


The answer to the question of whether the disk in BHBs becomes truncated in the luminous hard state has been controversial for several years, and \gx339, as an archetypical BHB, has been extensively studied with reflection spectroscopy (e.g., \citealp{tomsick2008broadband, petrucci2014return, plant2015truncated, basak2016spectral, jiang2019high}). Early observations of the disk truncation radius in the luminous hard state were controversial because pile-up could affect the shape of the iron line \citep{miller2006long, reis2008systematic,done2010re}. Recent analysis with \rxte\ \citep{javier_gx339} and \nustar\ \citep{wang2018evolution} has allowed for a more reliable determination of the reflection spectrum, and suggested the disk truncation level was below $\sim10R_g$ ($R_g=GM_{\rm BH}/c^2$) when $L>1\%L_{\rm Edd}$. However, debate still ensues about the choice of reflection model and underlying continuum \citep{mahmoud2019reverberation, dzielak2019comparison}. 




With reflection spectroscopy focusing on the time-integrated energy spectra, Fourier timing techniques have been developed more recently, which could quantify the multi-timescale variability and the corresponding time delays between energy bands (see \citealp{uttley2014x} for a review). The reverberation signal is the time lag introduced by light-travel time differences between observed variations in the direct power-law and the corresponding changes in the reflection spectrum. The first significant reverberation lag was detected in the AGN 1H 0707--495 \citep{fabian2009broad}, with soft-excess emission lagging the continuum-dominated band above $7\times10^{-4}$ Hz (equivalent to timescales shorter than 30 min). Later, other reverberation signatures, lags at the iron K emission line at $\sim6.4$ keV, were revealed by lag-energy spectra \citep{zoghbi2012relativistic, kara2016global}, which follow self-consistently the reflection picture. For BHBs, the detection of reverberation lags is more difficult because the number of received photons per light-travel time (determined by $R_g$, which is different by a factor of $10^5$ or more) is much smaller. Disk thermal reverberation lags were first detected for GX 339--4 \citep{uttley2011causal, de2015tracing}, and later also in H1743-322 \citep{de2016reverberation}. 

High frequency soft lags have been interpreted as due to reverberation. However, directly converting the amplitude of the time delays to a light travel distance has led to suggestions that the corona-disk distance could be hundreds of $R_g$, whereas X-ray reflection spectroscopy suggests small disk truncation of $<10R_g$ (e.g. \citealt{javier_gx339}). This highlights another aspect of the truncation debate in the hard state. Therefore, with \nicer's superior time resolution and no pile-up, we want to measure the reverberation lag, and compare to energy spectral fitting and the lag measured with \xmm\ data \citep{de2015tracing, de2017evolution}. 


Besides reverberation signatures, it has long been known that there are fairly ubiquitous hard lags at low temporal frequencies (below $\sim300 M_\odot/M$ Hz), both in AGN \citep{papadakis2001frequency, mchardy2004combined, arevalo2006investigating, mchardy2007discovery} and BHBs \citep{van1987complex, miyamoto1989x, vaughan1994time, nowak1999rossi}. These hard lags are difficult to explain with models invoking light-travel time delays. For instance, Compton upscattering of photons in the corona would require a corona that is extremely large to match the large amplitude of the hard lags \citep{nowak1999rossi}; the reverberation delay of reflection from a disk cannot explain the sign of the lags \citep{cassatella2012joint}. Therefore, a non-reverberation explanation is needed, which is also consistent with the fact that the corresponding lag-energy spectra only display a featureless energy dependence, without any reflection features. The most promising interpretation at present is propagating mass accretion rate fluctuation first proposed by \citet{lyubarskii1997flicker}.




This paper is organized as follows. Section~2 describes the observations and data reduction, Section~3 provides the details of energy spectral fitting. We present the time lag analysis with \nicer\ data in Section~4, discussion in Section~5, and summarize the results in Section~6. 

\section{Observations and data reduction}
\begin{figure}[htb!]
\centering
\includegraphics[width=\linewidth]{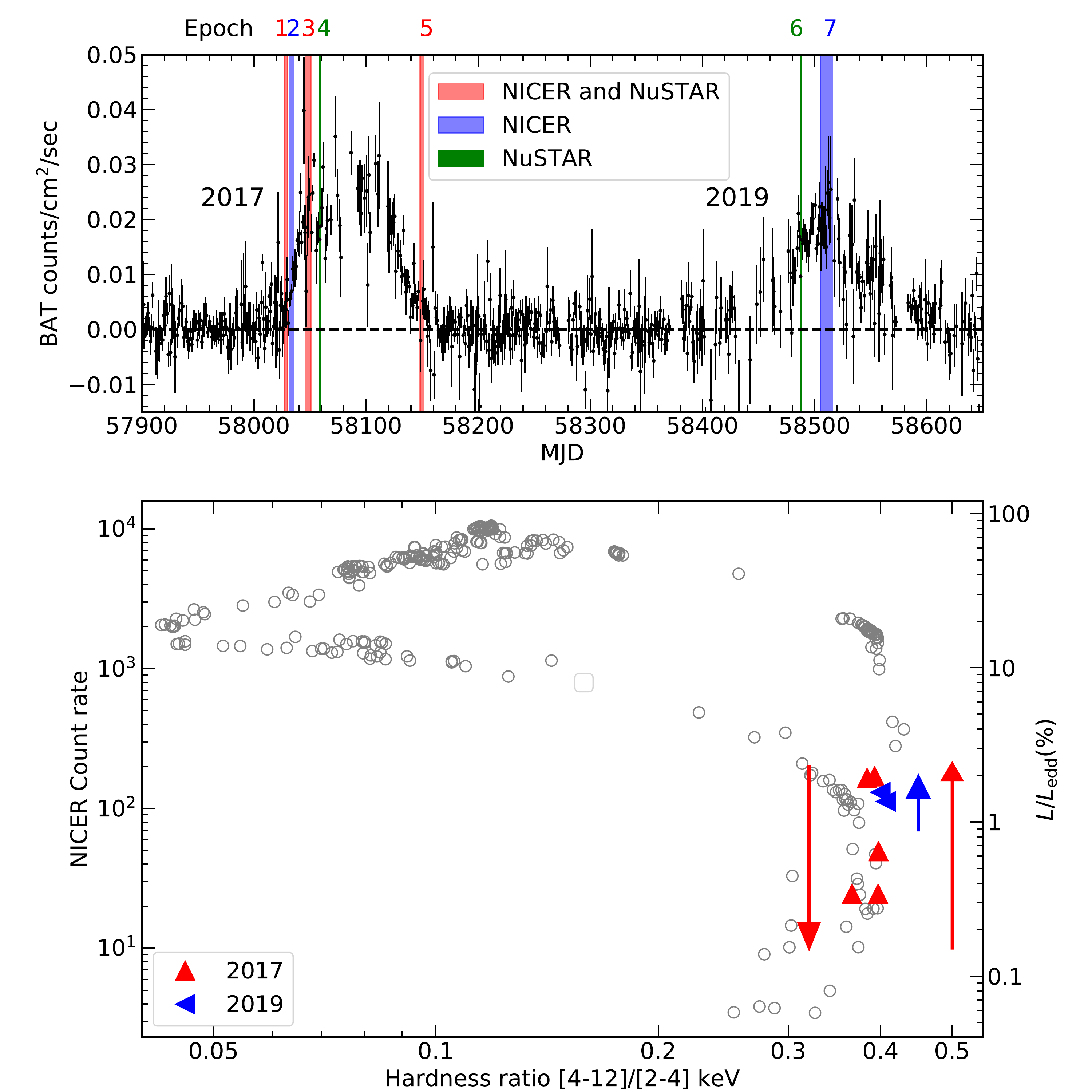}
\caption{
(\textit{Upper}) The \swift\ /BAT light curve (15-50~keV) of \gx339\ during 2017 and 2019 hard-only outbursts, with 7 epochs we chose labeled using vertical shaded regions. Different color codings represent the data availability in each epoch. (\textit{Lower}) The \nicer\ hardness-intensity diagram (HID) where the hardness ratio is defined as the count ratio of the hard band (4-12~keV) and the soft band (2-4~keV). The grey circles are simulated from RXTE PCU-2 spectral data in the 2002-2003 outburst, by first fitting  with a Comptonized disk-blackbody model from 3-45~keV.  For each data set, the best-fitting model was then convolved with the \nicer\ response matrix to produce the simulated HID track which is shown.
}
\label{fig:lc}
\end{figure}

\gx339\ entered an outburst in September 2017 after an optical brightening \citep{russell2017new}. Then, multi-wavelength observations were triggered, showing a flux rise in both radio \citep{russell2017radio} and X-ray \citep{gandhi2017x}. \nicer\ monitored \gx339\ on a 1-2 days cadence from 2017 September 29 to October 23, and 2018 January 23 to February 26, with the gap caused by solar angle constraints. In 2019, \nicer\ made observations from Jan 22 to Feb 2, catching the peak of another hard-only outburst (see Epoch~7 in Fig.\ref{fig:lc}). The relevant ObsIDs are 1133010101 through 1133010147. 

We adopt the same 2017 \nustar\ dataset including 4 observations (ObsIDs 80302304002 through 80302304007) as in \citet{garcia20192017}, in which simultaneous \swift\ observations are used to cover the soft energy band. As for the 2019 outburst, \nustar\ made two Target of Opportunity observations. Only the latter one on January 5 2019 is used (ObsID 90401369004) because of the short exposure time of the other ($<1$~ks). The \swift/BAT \citep{krimm2013swift} light curve and the \nicer\ HID are shown in Fig.\ref{fig:lc}, where the background grey HID track is simulated from \rxte\ spectra in \gx339's 2002-2003 outburst.

The \nicer\ data are processed with \nicer\ data-analysis software (NICERDAS) version 2017-09-06\_V002 and CALDB version 20170814. We use the standard filtering criteria: the pointing offset is less than $54\arcsec$, the pointing direction is more than $40^\circ$ away from the bright Earth limb, more than $30^\circ$ away from the dark Earth limb, and outside the South Atlantic Anomaly (SAA). In addition, we select events that are not flagged as ``overshoot" or ``undershoot" resets (EVENT\_FLAGS=bxxxx00), or forced triggers (EVENT\_FLAGS=bx1x000). A ``trumpet" filter is also applied to filter out known background events \citep{bogdanov2019constraining}. We select \nicer\ observations with at least one good time interval (GTI) longer than 60~s, and extracted individual spectra for each GTI. The cleaned events are barycenter corrected using the \texttt{FTOOL} \texttt{barycorr}. Our data sets were comprised of observations taken during observatory day and night times. The day-time data gains were corrected for optical loading due to a light leak on the instrument. Calibration uncertainties from these were reduced by correcting the spectra using residuals of a power-law fit to the Crab Nebula, a method referred to as ``Crab Correction" \citep{ludlam2018detection}. The background spectra are obtained using \nicer\ background model 3C50\_RGv5 (Remillard et al. in prep). To boost signal-to-noise ratios, we combine close-in-time individual spectra with similar hardness ratios ([4-12]/[2-4]~keV) and count rates (see Table~\ref{tab:obs}). The spectra are then binned with a minimum count of 1 per channel, and the oversampling factor is 3. The fitted energy range is 0.4-10~keV because the spectra become background-dominated above 10~keV. The response matrices we use in spectral fitting are nicer\_v1.02.rmf and ni\_xrcall\_onaxis\_v1.02.arf. For the purpose of timing analysis, the light curve segment length is 10~s with 0.001~s bins, which covers frequencies from 0.1 to 500~Hz. 


\nustar\ data are reduced using the Data Analysis Software (NUSTARDAS) 1.8.0 and CALDB v20170817. Source spectra are extracted from $60\arcsec$ circular extraction regions centered on the source position, and background spectra from $100\arcsec$ off-source circular regions. The \nustar\ data taken during 2019 (ObsID 90401369004) is only available in observation mode 06 because the source was Sun constrained, so a specific NUSTARDAS software module ``nusplitsc" is used to generate event files for different combinations of the three star tracker camera units. The source is extracted from $120\arcsec$ circular regions, background from $100\arcsec$ off-source circular regions. The spectra are then generated by the standard ``nuproducts" task. The spectra for different combinations of the three star tracker camera units in ObsID 90401369004 are combined using the FTOOL ``addspec". The FPMA/B spectra are binned with a minimum signal-to-noise of 5 per bin, and the oversampling factor was 3. The fitted energy range is 3-79~keV. 

With the availability of data, we choose 7 epochs in total to cover the two hard-only outbursts (see Fig.~\ref{fig:lc} and Tab.~\ref{tab:obs}). Epochs~1-4 were taken during the rising phase of the 2017 outburst, reaching a maximal $1-100$~keV unabsorbed luminosity of $2.1\%$~$L_{\rm Edd}$ assuming a distance of 8 kpc and a black hole mass of
$10~M_{\odot}$, while Epoch~5 was in the decay phase, with a luminosity similar to Epoch~1. Epochs~6 and 7 were both in the rising phase of the 2019 outburst, and the peak luminosity ($1.6\%$~$L_{\rm Edd}$) is slightly lower than the peak in 2017. 

All the uncertainties quoted in this paper are for a 90\%
confidence range, unless otherwise stated. All spectral fitting is done with XSPEC 12.10.1f \citep{arn96}. In all of the fits, we use the \textit{wilm} set of abundances \citep{wilms2000}, and \textit{vern} photoelectric cross sections \citep{verner1996atomic}. The fitting statistics is PG-statistics\footnote{\url{https://heasarc.gsfc.nasa.gov/xanadu/xspec/manual/XSappendixStatistics.html}}, for Poisson data with a Gaussian background.

\begin{table*}[htb!]
\begin{center}
\caption{\nicer\ and \nustar\ observations in the 2017 and 2019 outburst
cycles, exposure times and dates. \label{tab:obs}}
\footnotesize
\begin{tabular}{ccc|cccc|cccc}\hline \hline
outburst&Epoch&$L/L_{edd}$&\multicolumn{4}{c|}{\textit{NICER}}&\multicolumn{4}{c}{\textit{NuSTAR}}\\
&&(\%)&obs.ID$^\dagger$&Date&exp.(ks)&counts/s&obs.ID&Date&exp.(ks)&counts/s\\
\hline
2017&1&0.3&03-05&10/01-10/04&7.8&24.5&80302304002&10/02&21.5&2.0\\
&2&0.6&07-10&10/06-10/09&5.0&49.4&-&-&-&-\\
&3&1.8&12-15&10/20-10/23&7.4&164.6&80302304004&10/25&18.0&22.8\\
&4&2.1&-&-&-&-&80302304005&11/02&18.9&16.9\\
&5&0.3&27-29&01/31-02/02 2018&3.9&24.5&80302304007&01/30 2018&29.0&3.2\\
\hline
2019&6&1.4&-&-&-&-&90401369004&01/05&3.6&14.4\\
&7&1.6&39-47&01/22-02/02&19.3&130.6&-&-&-&-\\
\hline
\end{tabular}
\\
\raggedright{\textbf{Notes.} \\Luminosity is calculated using the unabsorbed flux
between $1-100$~keV, assuming a distance of 8 kpc and a black hole mass of
$10~M_{\odot}$. The count rates are for the fitted energy ranges, i.e., $0.4-10$~keV for \textit{NICER} and $3-79$~keV for \textit{NuSTAR}. $^\dagger$ObsIDs for \nicer\ are 11330101xx. }
\end{center}
\end{table*}

\section{Spectral Fitting} \label{fit}

\subsection{Towards the final model} \label{towards_model}
\begin{figure*}[htb!]
\centering
\includegraphics[width=1.\linewidth]{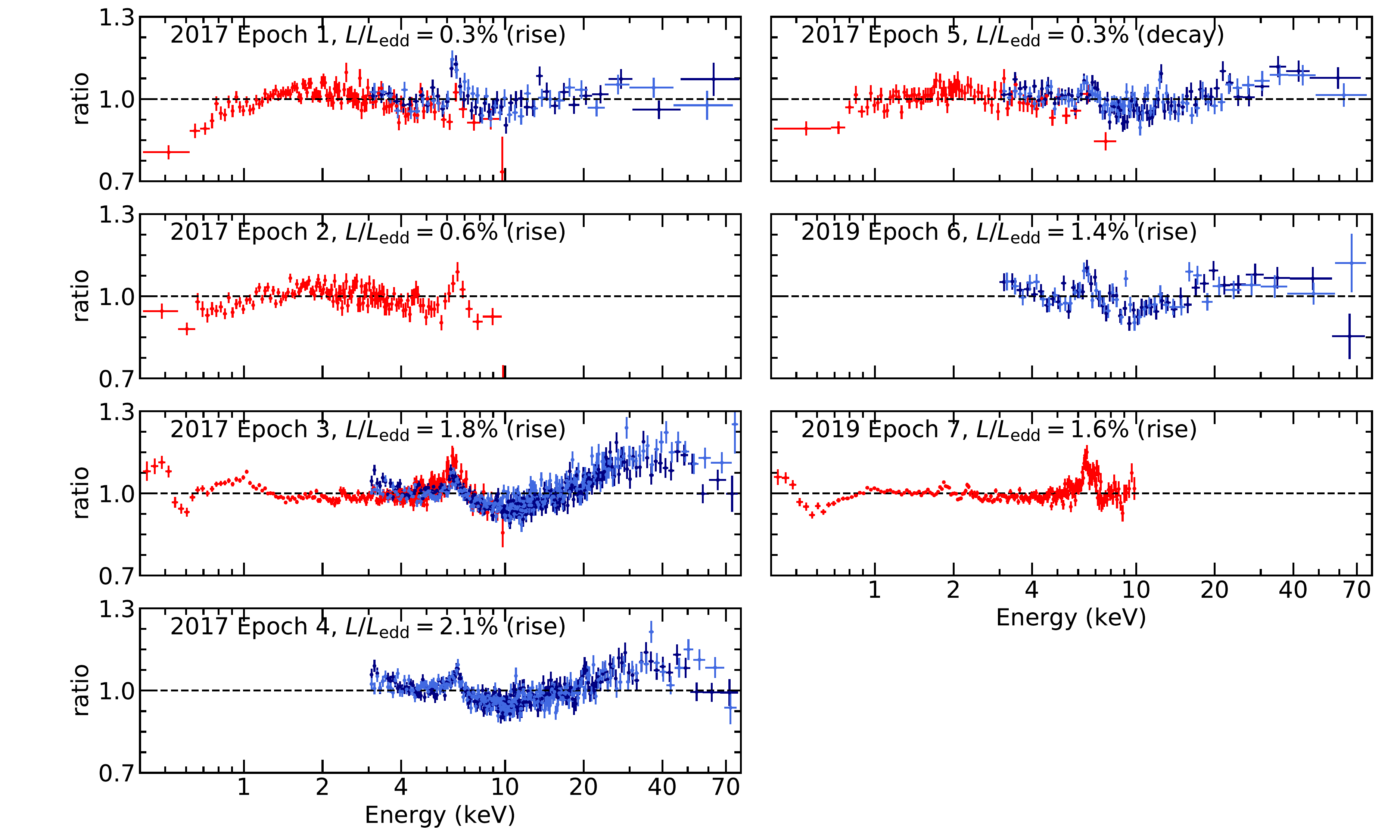}
\caption{Data-to-model ratio for the simultaneous fit with absorbed power-law model (\texttt{Tbabs*crabcorr*powerlaw}). The iron line complex around 6.4~keV are prominent in all epochs, and Compton hump above 20~keV are also present in Epoch~3 and 4 with the highest luminosities. Red: \nicer\ , navy: \nustar\ /FPMA, blue: \nustar\ /FPMB. 
}
\label{fig:ratio_po}
\end{figure*}

In order to assess the reflection features, we first fit the 15 spectra (5 \nicer\ spectra and 10 \nustar\ spectra accounting for both FPMA and FPMB) simultaneously with an absorbed power-law model (\texttt{Tbabs*crabcorr*powerlaw} in \xspec\ notation). The cross-calibration between \nicer\ and \nustar\ is carried out by the model \texttt{crabcorr} \citep{steiner2010constant}, which could multiply each model spectrum by a power-law, applying corrections to both the slope of power-law via the parameter $\Delta\Gamma$, and normalization. In this way, the responses of different detectors are cross-calibrated to return the same normalizations and power-law slopes for the Crab. The column density $N_H$ is tied, while photon-index $\Gamma$ is free to vary between epochs. The resulting data-to-model ratios are shown in Fig.~\ref{fig:ratio_po}, where the iron line complex is prominent in all epochs, and the Compton hump can also be seen in some of the spectra. This fit gives a Galactic column density of $N_H\sim5.8\times10^{21}$~cm$^{-2}$.

Next, we perform a simultaneous fit to all 7 epochs using the relativistic reflection model \texttt{relxillCp}. In \xspec\ notation, the model is \texttt{Tbabs*crabcorr*relxillCp}. The reflection model \texttt{relxillCp} includes both the original continuum emission from the ``corona" and the reprocessed emission from the disk. The coronal emission is described by \texttt{nthcomp} with the seed photons originating from the disk \citep{zdziarski1996broad,zycki19991989}. The spin $a_*$ has been measured to be high for \gx339 (e.g., $>0.97$ in \citealt{ludlam2015reapproaching}, and $0.95^{+0.02}_{-0.08}$ in \citealt{parker2016nustar}). Therefore, the spin parameter is fixed at the maximal value of 0.998 to allow the inner disk radius to fit to any physically allowed value. We find that the electron temperatures in the corona ($kT_{\rm e}$) can only be constrained to have lower limits, i.e., pegged at the maximal value, so we fix $kT_{\rm e}=400$~keV in the subsequent fittings; we also confirm that allowing $kT_{\rm e}$ to vary does not improve the fit ($\Delta$PG-stat $<1$), and other parameter values stay the same. With regard to the emissivity profile, we choose the canonical profile of $\propto r^{-3}$ (emissivity index $q=3$), and we explore these effects in Section~\ref{other_emissivity}. This fit results in PG-statistics/d.o.f.=7702/5827=1.32. 

\begin{figure}[htb!]
\centering
\includegraphics[width=1.\linewidth]{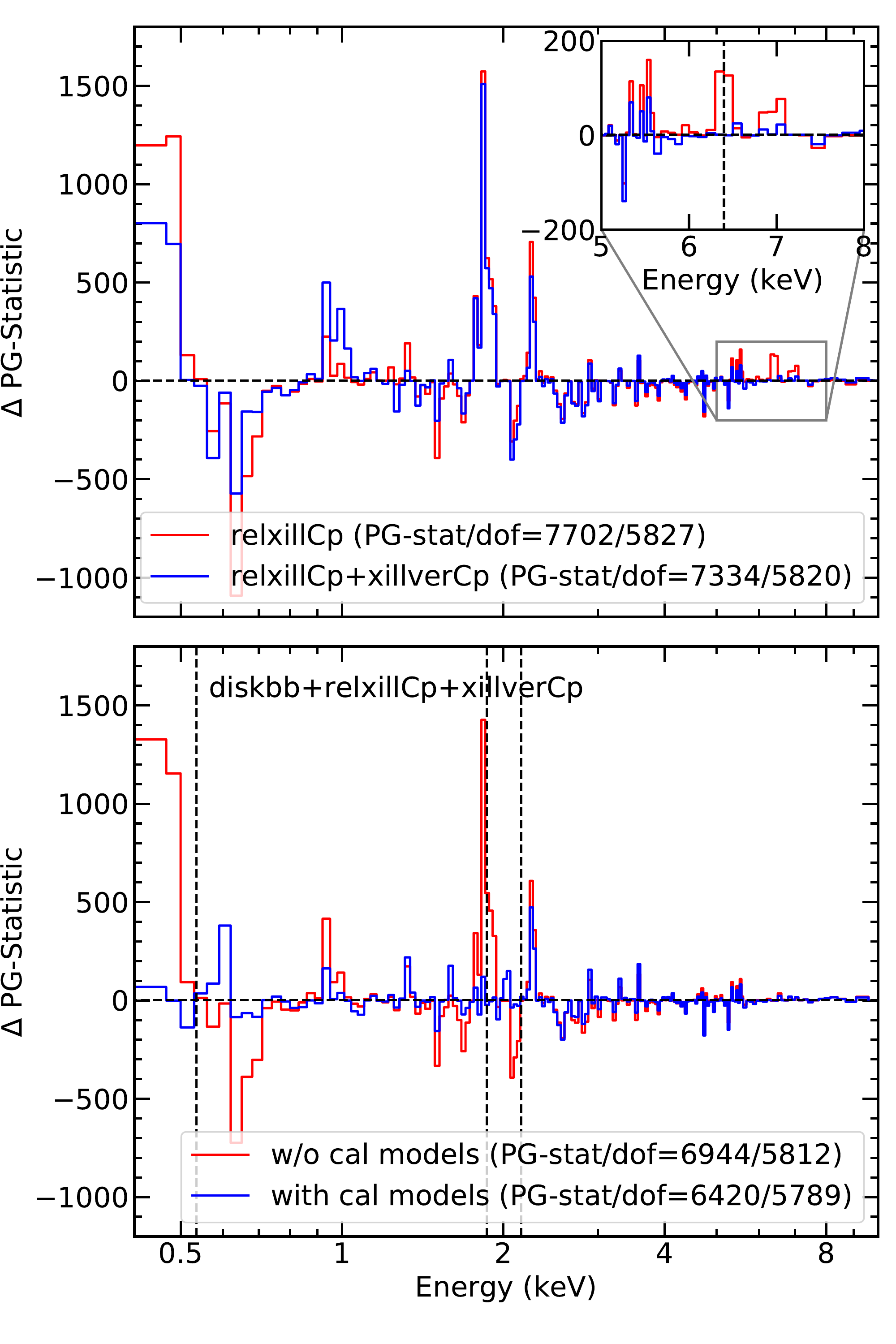}
\caption{Residuals for \nicer\ data in Epoch 7 from simultaneous fits with different models. (\textit{Upper}) Reflection from a distant reflector is needed to fully describe the iron line region. (\textit{Lower}) red: without the phenomenological calibration models, i.e. \texttt{Tbabs*crabcorr*[model shown]}); blue: with the calibration models, i.e. \texttt{Tbabs*crabcorr*([model shown]+gaussian)*edge*edge}. The best-fitted energies in calibration models are also plotted with vertical dashed lines. Notice that the PG-statistics values are for simultaneous fits, even though only the \nicer\ data in Epoch 7 are shown. }
\label{fig:series_mo}
\end{figure}

However, similar to earlier works on \gx339\ \citep{javier_gx339, wang2018evolution, jiang2019high}, we find that a single component of relativistic reflection could not fully describe the iron line region (see the upper panel in Fig.~\ref{fig:series_mo} showing the zoom-in residuals around the iron line of \nicer\ data in Epoch~7 as an example), and that an extra distant reflector modeled by \texttt{xillverCp} could solve this problem. Within \texttt{xillverCp}, parameters describing the properties of the corona (photon index $\Gamma$, electron temperature $kT_{\rm e}$) and the system (inclination $i$, Fe abundance $A_{\rm Fe}$) are tied to those in \texttt{relxillCp}. Also, the reflector is assumed to be close to neutral, i.e., $\log \xi$~(erg$\cdot$cm$\cdot$s$^{-1}$) $=0$. We find that if $\log\xi$ is free to vary, PG-stat decreases by 11 with 7 fewer parameters, which is not significant; the constraints on $\log\xi$ are loose, and the other parameter values are not affected. These setups only add one extra free parameter per epoch, namely the normalization of \texttt{xillverCp}. With this model (\texttt{Tbabs*crabcorr*(relxillCp+xillverCp)}), the PG-statistics reduces by 368 with 7 fewer d.o.f., and the residuals around the iron line are largely diminished as expected. 

As shown in Fig.~\ref{fig:series_mo} (\textit{upper}), strong features at soft energies ($<3$~keV) are still present. A straightforward explanation is a thermal disk component, so we try the model of \texttt{crabcorr*Tbabs*(diskbb+relxillCp+xillverCp)}. Since the disk component is not visible in \nustar 's energy range, the disk component in Epochs~4 and 6 which consist of \nustar\ data alone are tied to those in Epochs~3 and 7 respectively. For the other 5 epochs, disk temperatures in Epoch~1 and 5 with the lowest luminosities ($\sim0.3\%$~$L_{\rm Edd}$) are pegged at the lowest value allowed, so we fix the disk temperature at a reasonably low value of 50~eV in these two epochs, and only let the normalizations free, to obtain a putative estimate of the unscattered flux from the disk component, i.e., disk photons which are not scattered when passing through the corona; while in Epoch~2, 3, and 7 with luminosities above $0.5\%$~$L_{\rm Edd}$,  the disk temperature is determined to be $100-200$~eV based on \nicer\ data. This model further reduces PG-statistics by 390 with 8 extra free parameters, resulting in PG-statistics/d.o.f.=6944/5812=1.19. 

\begin{figure*}[htb!]
\centering
\includegraphics[width=.95\linewidth]{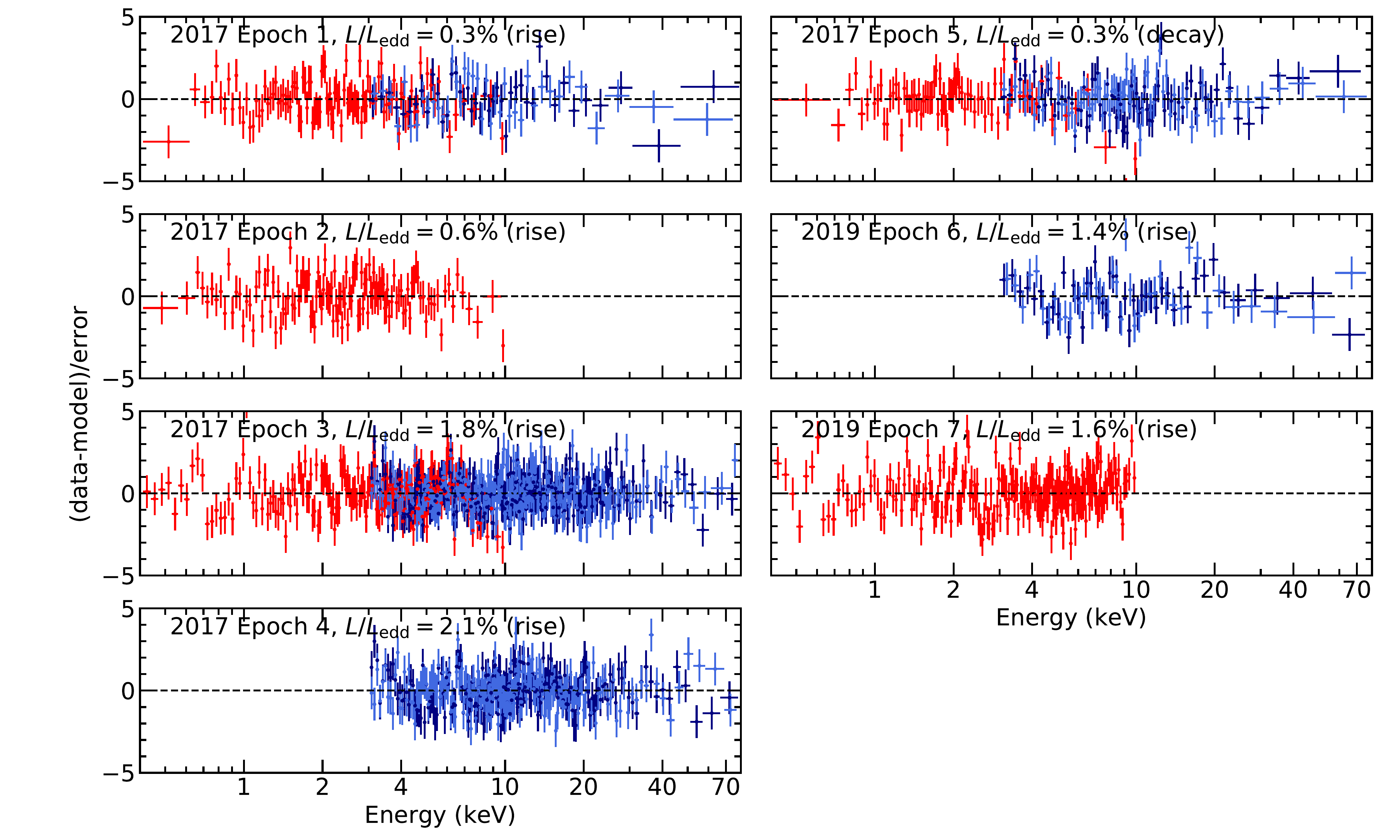}
\includegraphics[width=.95\linewidth]{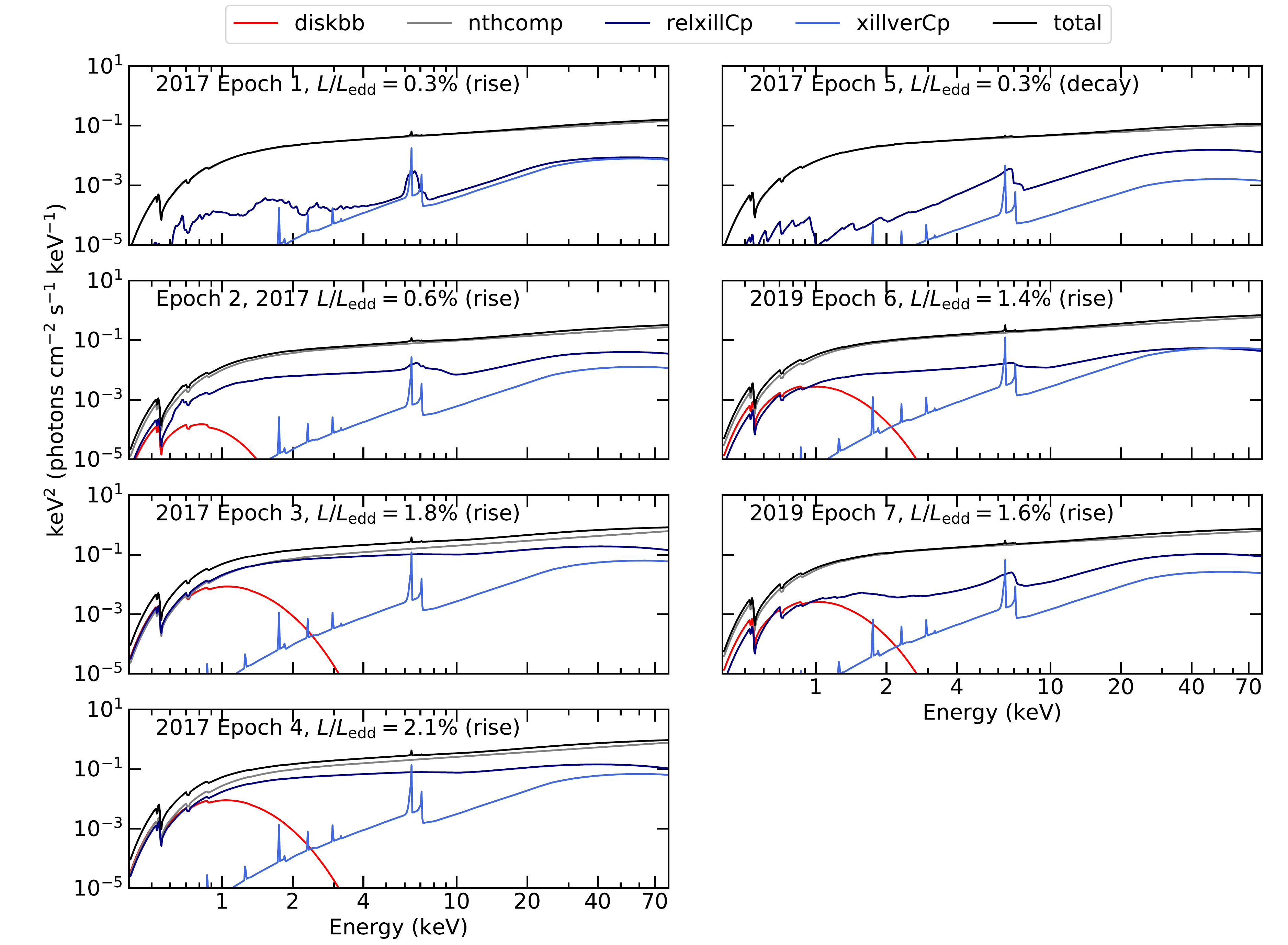}
\caption{(Data-model)/error (\textit{upper}) and model components (\textit{lower}) for the simultaneous fit with the final complete model (\texttt{Tbabs*crabcorr*(diskbb+relxillCp+xillverCp+gaussian)*edge*edge}). In the ratio plot, red: \nicer\ , navy: \nustar\ /FPMA, blue: \nustar\ /FPMB. Notice that the continuum component via \texttt{nthcomp} that is included in the \texttt{relxillCp} model is plotted separately from the reflection component. Also, the \texttt{gaussian} component that accounts for the calibration uncertainty is not shown, and all the information about the calibration model is present in the Table~\ref{tab:parameters}. 
}
\label{fig:ratio_final}
\end{figure*}

We notice that there are still some large residuals at soft energies (see Fig.~\ref{fig:series_mo} lower). The most noticeable features include edge-like shapes near $\sim0.5$~keV and $\sim2.2$~keV, and Gaussian-like emission around $1.8$~keV. These energies correspond to features in \nicer's effective area versus energy, where 0.5, 1.8, and 2.2~keV features are attributed to oxygen, silicon, and gold, respectively. Therefore, we have reason to believe that these remaining features come from \nicer's calibration systematics, and we model them empirically in our work. We suggest that observers be aware of such residuals and follow the latest analysis guidelines of the NICER analysis. The final model we adopt can be expressed as follows

\begin{equation}
\begin{multlined}
	\texttt{crabcorr*Tbabs*(diskbb+relxillCp+xillverCp}\\ \texttt{+gaussian)*edge*edge} \nonumber
\end{multlined}
\end{equation}

where the \texttt{gaussian} and the two \texttt{edge} components are phenomenological models to account for residual calibration features. In each model, the energy is tied through the 5 epochs containing \nicer\ data, while the other quantities ($\sigma$ and norm in \texttt{gaussian}, $\tau_{\rm max}$ in \texttt{edge}) can vary. This allows for potential attitude dependent variations, which are not accounted for in the current response functions. In comparison with the fit without calibration models, the PG-stat/d.o.f. becomes 6420/5789=1.11, decreasing by 524 with 23 fewer d.o.f. We also emphasize that the parameters coming out of the reflection modeling are not changed when accounting for the calibration features. The residuals of \nicer\ data in Epoch~7 for these two fits are shown in Fig.~\ref{fig:series_mo} (\textit{lower}), where the best-fitted energies in calibration models are also plotted with vertical dashed lines. The (data-model)/error for all the epochs, and the model components are shown in the upper and lower portions of Fig.~\ref{fig:ratio_final}. The best-fit parameters are presented in Table~\ref{tab:parameters}. 

\subsection{Global parameters} \label{global_par}
In this simultaneous fit with $\sim5.2$~million counts, the column density (in units of $10^{21}$cm$^{-2}$) is constrained to be $N_{\rm H}=6.41^{+0.08}_{-0.09}$, which is consistent with previous X-ray reflection spectroscopy results including, e.g., the value determined by \suzaku\ ($\sim6.8$ in \citealt{tomsick2009truncation}, $4.7-6.7$ in \citealt{petrucci2014return}), \xmm\ ($\sim7.0$ in \citealt{basak2016spectral}), and \swift\ ($4.1-6.9$, \citealt{wang2018evolution}). From optical reddening, $N_{\rm H}$ is $6.0\pm0.6$ \citep{zdziarski1998broad}, also consistent with our result. 

We find that the inclination angle is ($38^{+2}_{-3}$)$^\circ$, consistent with $\sim40^\circ$ in \citet{wang2018evolution}, and slightly smaller than $\sim48^\circ$ in  \citet{javier_gx339}. Previous reflection spectroscopy studies have constrained the inclination to $i=30-60$$^\circ$, with the exact value being model-dependent \citep{javier_gx339, steiner2017self}. Also, our result is in agreement with the inclination found with \suzaku\ data ($36\pm4$$^\circ$, \citealt{ludlam2015reapproaching}). In addition, the latest ellipsoidal light curve in the NIR band has shown that the binary inclination is $37^\circ<i<78^\circ$ \citep{heida2017potential}, whose lower limit is close to our result here. The inclination obtained with reflection spectroscopy is for the inner accretion disk, and could possibly be different from the binary inclination. 

Another global parameter in our simultaneous fit is the iron abundance, found to be $A_{\rm Fe}=4.08^{+0.15}_{-0.22}$ in solar unit. This is still super-solar, and as shown in \citet{javier_gx339,wang2018evolution}, we confirm that this preference comes from a significant reduction of residuals at high energies seen by \nustar . A fixed solar iron abundance ($A_{\rm Fe}=1$) would increase PG-stat by 276. We note that the super-solar iron abundance problem could be potentially solved by adopting the high-density reflection model, as explored in \citet{tomsick2018alternative} and \citet{jiang2019high}. Especially, \citet{jiang2019high} analyzed the same dataset as in \citet{wang2018evolution}, and obtained a close-to-solar iron abundance of $1.50^{+0.12}_{-0.04}$~$A_{\rm Fe, \odot}$ and a high density in the disk surface of $\log(n_e/{\rm cm}^{-3}) = 18.93^{+0.12}_{-0.16}$; while the density is fixed at $\log(n_e/{\rm cm}^{-3}) = 15$ in the reflection model adopted in this work. We tested the high density reflection model for Epoch~3, but found an increased $A_{\rm Fe}$. More details can be found in the Appendix.

\subsection{Evolution of physical properties} \label{evolution}

\begin{figure}[htb!]
\centering
\includegraphics[width=1.\linewidth]{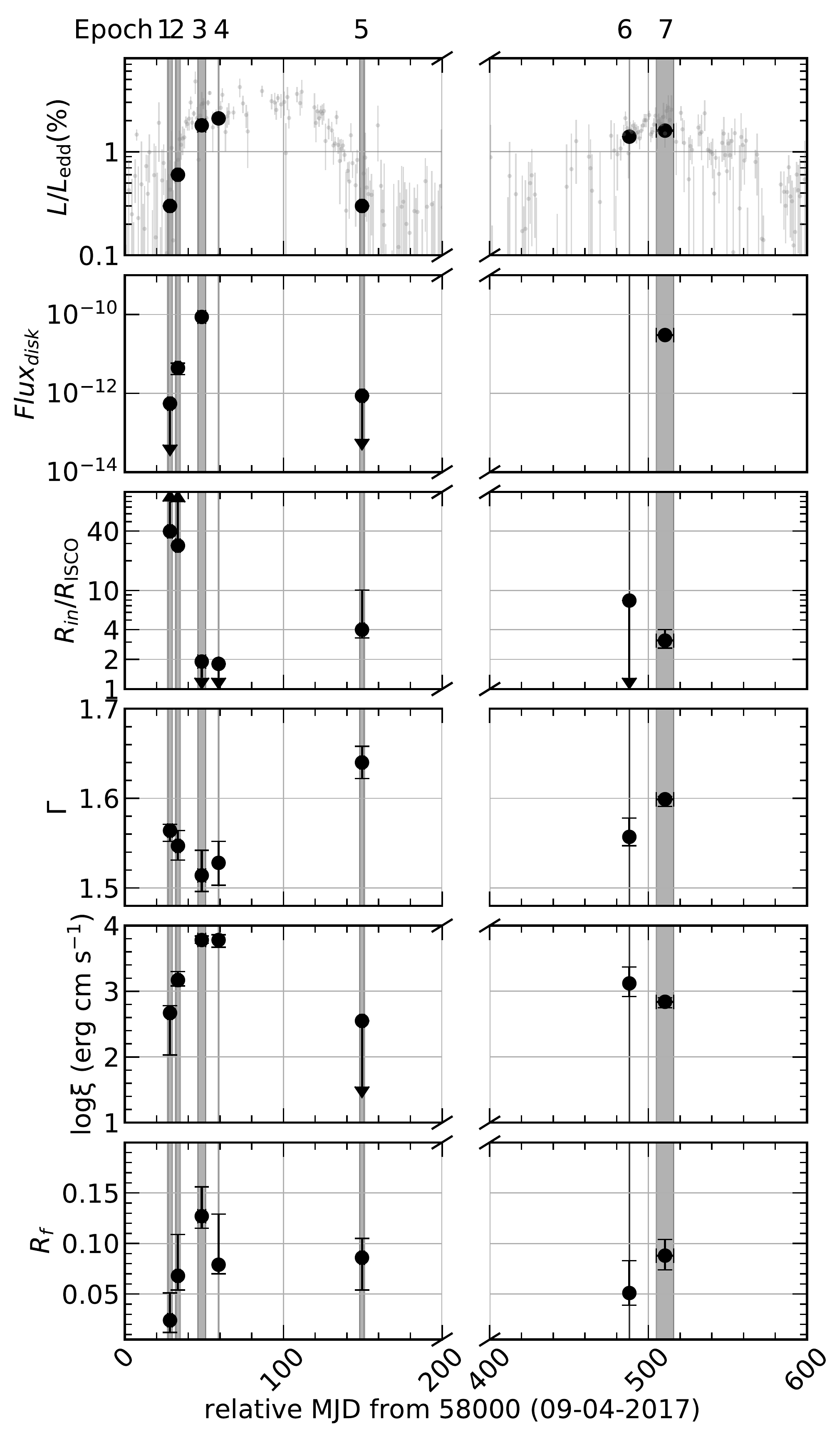}
\caption{The best fit evolution of disk properties during the 2017 and 2019 outburst cycles. Each epoch is shown with vertical shaded regions. In the top panel, the \swift /BAT light curve is overplotted to suggest the outbursts. The exact parameter values can be referred to Table~\ref{tab:parameters}. 
}
\label{fig:par_evol}
\end{figure}

In addition to these values of global parameters, we see clear evolution in the properties of the disk (see Fig.~\ref{fig:par_evol}). First, we examine its evolution during the rise in 2017 (Epoch~1--4), when the luminosity increases from 0.3\% to 2.1\% $L_{\rm Edd}$. 

\begin{itemize}
	\item The inner radius of the disk only has a lower limit in Epoch~1 ($>40$ innermost stable circular orbit radius, $R_{\rm ISCO}$ afterwards) and Epoch~2 ($>28R_{\rm ISCO}$), suggesting a large truncation radius of the disk, while in Epoch 3 and 4 with luminosities $\ge1.8\%L_{\rm Edd}$, $R_{\rm in}<2~R_{\rm ISCO}$. This trend is in line with the commonly agreed picture that the inner disk moves inwards as the luminosity increases \citep{esin1997advection, meyer2005hysteresis, kylafis2015accretion, marcel2019unified}. 
	\item The unabsorbed and unscattered flux from the disk component calculated using the model \texttt{cflux} in \xspec\ increases from $<5.5\times10^{-13}{\rm ergs}/{\rm cm}^2/{\rm s}$ to $\sim 8.7\times10^{-11}{\rm ergs}/{\rm cm}^2/{\rm s}$, by more than 2 orders of magnitude. In the meantime, the disk temperature is fixed at 50~eV in Epoch~1, and becomes $\sim100$~eV and $\sim200$~eV in Epoch~2 and 3 respectively. 
	
	\item The ionization parameter increases from $\log \xi$~(erg$\cdot$cm$\cdot$s$^{-1}$) $=2.7$ to 3.8. 
	
	\item The reflection fraction $R_f$, which describes the fraction of reflected photons to those reaching the observer directly \citep{dauser2016normalizing}, also increases from Epoch~1 (0.02) to 3 (0.13). This could be explained by a decreasing inner radius of the disk. In Epoch~4, the $R_f$ is $0.079^{+0.050}_{-0.009}$; while the best-fit value is not as large as in Epoch~3, its 90\% confidence upper limit is 0.13, consistent with $R_f$ in Epoch~3. 
	
	
\end{itemize}

Because of a larger effective area and longer exposure times, the \nicer\ spectra have much better signal-to-noise ratios than \swift\ (which were used in \citealt{wang2018evolution} and where disk evolution was difficult to determine), we are now able to obtain reasonable and self-consistent evolution which are all predicted as the accretion disk's inner radius moves inwards between 0.3\% to 2.1\%~$L_{\rm Edd}$. One exception is the photon index $\Gamma$ which remains quite constant, and we will discuss this in Section~5.1. 

Moreover, it is also worthwhile to compare results at the beginning and the end of the 2017 outburst (Epoch~1 and 5), since as mentioned earlier, the determined luminosities are similar, and both epochs have \nicer\ and \nustar\ data for a broader energy coverage, while taken during the rise and decay in the 2017 hard-only outburst. With the disk temperature fixed at 50~eV, the flux from the disk component during each epoch have a similar putative limit. The ionization has only an upper limit, $\log \xi$~(erg$\cdot$cm$\cdot$s$^{-1}$) $<2.55$. Also, the spectrum is slightly softer in the decay, with $\Gamma\sim1.64$ compared to $\sim1.56$. This could be attributed to a difference in the optical depth in the corona. The most interesting difference is that during the decay, $R_{\rm in}=4.0^{+6.1}_{-0.7}$~$R_{\rm ISCO}$, while during the rise, the disk is largely truncated ($R_{\rm in}>40R_{\rm ISCO}$). This difference naturally accounts for a reflection fraction $\sim$4 times larger in Epoch~5 than in Epoch~1. The different level of disk truncation suggests a hysteresis effect during the rise and decay. It is worth noticing that if the inner radius in Epochs~1 and 5 are tied together, the PG-stat increases by 8, and $R_{\rm in}>36R_{\rm ISCO}$. The confidence contours of $R_{\rm in, 1}$ and $R_{\rm in, 5}$ in the simultaneous fit suggest that they differ at the 2-$\sigma$ level. While alone this is not statistically significant, we are encouraged by the simultaneous change in ionization and reflection fraction, all consistent with the hysteresis picture. If the hysteresis effect is real, then the lack of a strong thermal component in Epoch~5 when the disk is only slightly truncated could be due to a low mass accretion rate. 


In the 2019 hard-only outburst (Epoch~6 and 7 covered by \nustar\ and \nicer\ respectively), \gx339\ has reached a lower peak luminosity (1.6\% compared to 2.1\%~$L_{\rm Edd}$). The inner disk radius is $\sim3R_{\rm ISCO}$, larger than in 2017. Meanwhile, the disk temperature ($\sim180$~eV), and unscattered flux from the disk component ($\sim3\times10^{-11}{\rm cm}^{-3}$), are both slightly lower; the disk is less ionized, and the reflection fraction is also lower. Otherwise, we observe no significant differences between the 2017 and 2019 hard-only outbursts, considering the peak luminosity difference.

\begin{table*}[htb!]
\caption{Best fit parameters for the final model \texttt{Tbabs*crabcorr*(diskbb+relxillCp+xillverCp+gaussian)*edge*edge}.  \label{tab:parameters}}
{\renewcommand{\arraystretch}{1.3}%
\resizebox{\linewidth}{!}{\begin{tabular}{cccccccccc}\hline \hline
Model & Parameter & Epoch 1 & Epoch 2 & Epoch 3 & Epoch 4 &Epoch 5 &Epoch 6&Epoch 7\\
\hline
\textsc{Tbabs}&$N_H$ ($10^{21}$cm$^{-2}$)&  \multicolumn{7}{c}{$6.41^{+0.08}_{-0.09}$} \\
\textsc{relxillCp}&$a_*$& \multicolumn{7}{c}{$0.988$ (f)}  \\
\textsc{relxillCp}&$i$ (degrees)&  \multicolumn{7}{c}{$37.6^{+2.2}_{-2.9}$} \\
\textsc{relxillCp}&$A_{\rm Fe}$&  \multicolumn{7}{c}{$4.08^{+0.15}_{-0.22}$} \\
\textsc{relxillCp}&$kT_{\rm e}$ (keV)&\multicolumn{7}{c}{$400$ (f)} \\
\textsc{relxillCp}&$q$&\multicolumn{7}{c}{$3$ (f)} \\
\textsc{Gaussian}&$E$ (keV)&  \multicolumn{7}{c}{$1.861\pm0.011$} \\
\textsc{$\rm edge_1$} & $E$ (keV)&  \multicolumn{7}{c}{$2.161^{+0.025}_{-0.026}$} \\
\textsc{$\rm edge_2$}&$E$ (keV)&  \multicolumn{7}{c}{$0.535^{+0.004}_{-0.003}$} \\
\hline
\textsc{diskbb}&$T_{\rm in}$ (eV)& $50$ (f) & $106^{+38}_{-52}$ &  $193\pm5$& - & $50$ (f)& - & $178^{+20}_{-9}$  \\
\textsc{diskbb}& Norm ($10^3$)& $<600$ & $11^{+64}_{-9}$ &  $6.4^{+1.6}_{-1.2}$& - & $<950$& -  & $3.5^{+1.6}_{-1.1}$  \\
\textsc{diskbb}&$F_{\rm disk}$ ($10^{-13}$ ergs/cm$^2$/s)& $<5.5$ & $44\pm14$ & $867\pm12$& -  & $<8.7$ & - & $301\pm7$\\
\textsc{relxillCp}&$R_{\rm in}$ ($R_{\rm ISCO}$)& $>39.8$ & $>28.5$ & $<1.9$ & $<1.8$ & $4.0^{+6.1}_{-0.7}$ & $<7.9$ & $3.1^{+0.9}_{-0.5}$  \\
\textsc{relxillCp}&$\Gamma$& $1.564^{+0.007}_{-0.012}$ & $1.547^{+0.017}_{-0.016}$ & $1.514^{+0.028}_{-0.018}$ & $1.528^{+0.024}_{-0.025}$ & $1.640\pm0.018$ & $1.557^{+0.021}_{-0.010}$ & $1.599^{+0.003}_{-0.008}$\\
\textsc{relxillCp}&$\log \xi$ (erg$\cdot$cm$\cdot$s$^{-1}$) & $2.67^{+0.11}_{-0.64}$ & $3.17^{+0.13}_{-0.09}$ & $3.78^{+0.06}_{-0.05}$ & $3.78^{+0.08}_{-0.11}$ & $<2.55$ & $3.12^{+0.25}_{-0.20}$ & $2.84^{+0.06}_{-0.09}$ \\
\textsc{relxillCp}&$R_{\rm f}$& $0.024^{+0.027}_{-0.013}$ & $0.068^{+0.041}_{-0.014}$ & $0.127^{+0.029}_{-0.012}$ & $0.079^{+0.050}_{-0.009}$ & $0.086^{+0.019}_{-0.032}$ &  $0.051^{+0.032}_{-0.012}$ & $0.088^{+0.016}_{-0.014}$\\
\textsc{relxillCp}&Norm ($10^{-3}$)& $1.23\pm0.06$ & $2.31^{+0.10}_{-0.11}$ & $5.30^{+0.25}_{-0.50}$ & $6.54^{+0.20}_{-0.25}$ & $0.88\pm0.05$ & $5.0\pm0.3$ & $5.27^{+0.17}_{-0.07}$\\
\textsc{xillverCp}&Norm ($10^{-4}$)& $0.9^{+0.5}_{-0.6}$ & $1.4^{+1.0}_{-0.9}$ & $6.9\pm0.8$ & $7.5^{+1.0}_{-1.1}$ & $<0.5$ & $6.0^{+1.7}_{-2.9}$ & $2.9\pm0.7$\\
\textsc{Gaussian}& $\sigma$ (keV)& $>0.054$ & $>0.076$ & $>0.088$ & - & $>0.065$ & - & $0.059^{+0.019}_{-0.017}$ \\
\textsc{Gaussian}& Norm ($10^{-4}$) & $0.4\pm0.3$ & $1.1\pm0.6$ & $3.5\pm0.9$ & - &  $0.7^{+0.4}_{-0.5}$& -  & $3.2^{+0.6}_{-0.5}$ \\
\textsc{$\rm edge_1$}& $\tau_{\rm max}$ & $-0.030^{+0.020}_{-0.021}$ & $-0.024^{+0.019}_{-0.018}$ & $-0.018\pm0.009$ & - & $-0.053^{+0.029}_{-0.026}$ & - & $-0.026^{+0.006}_{-0.007ß}$ \\
\textsc{$\rm edge_2$}& $\tau_{\rm max}$ & $0.23^{+0.12}_{-0.06}$ & $0.46^{+0.11}_{-0.14}$ & $0.41^{+0.05}_{-0.06}$ & - &  $0.12^{+0.09}_{-0.08}$& - & $0.40\pm0.04$\\
\hline
\textsc{crabcorr}&$\Delta \Gamma_{\rm NICER}$& 0 (f) & 0 (f) & 0 (f)  & - & 0 (f) & - & 0 (f)\\
\textsc{crabcorr}&$N_{\rm NICER}$& 1 (f) & 1 (f) & 1 (f)  & - & 1 (f) & - & 1 (f)\\
\textsc{crabcorr}&$\Delta \Gamma_{\rm FPMA}$& $0.097^{+0.020}_{-0.015}$ & - & $-0.013^{+0.010}_{-0.008}$ & 0 (f) & $0.022^{+0.021}_{-0.019}$& 0 (f) & -\\
\textsc{crabcorr}&$N_{\rm FPMA}$& $0.93\pm0.03$ & - & $1.196^{+0.020}_{-0.019}$ & 1 (f) & $1.36\pm0.04$& 1 (f) & -\\
\textsc{crabcorr}&$\Delta \Gamma_{\rm FPMB}$& $0.102^{+0.020}_{-0.015}$ & - & $-0.039^{+0.011}_{-0.008}$ & $-0.001^{+0.005}_{-0.006}$ & $0.017\pm0.019$& $-0.012\pm0.016$ & -\\
\textsc{crabcorr}&$N_{\rm FPMB}$& $0.96\pm0.03$ & - & $1.160^{+0.022}_{-0.026}$ & $1.021\pm0.014$ & $1.40\pm0.04$& $1.01^{+0.03}_{-0.04}$ & -\\
\hline
&$L/L_{\rm Edd}$&0.3\% & 0.6\% & 1.8\% & 2.1\% & 0.3\% & 1.4\% & 1.6\% \\
&PG-Stat./d.o.f.& \multicolumn{7}{c}{$6420.5/5789=1.11$} \\
\hline
\end{tabular}}}
\\
\raggedright{\textbf{Notes.} \\
Errors are at 90\% confidence level and statistical only. The flux of the disk component is calculated using \texttt{cflux} in \xspec. For the silicon line modeled by \texttt{Gaussian}, the width is set to have an upper limit of 0.1~keV, and it is pegged at that upper limit in all epochs except for Epoch~7. }
\end{table*}

\subsection{Other emissivity profiles} \label{other_emissivity}
For the spectral fitting, we also tried emissivity profiles other than the canonical one that describes the intensity from the disk with $\propto r^{-3}$, including free emissivity indices and lamppost geometry. With emissivity indices free to vary, the PG-stat decreases by 34 with 7 fewer d.o.f., which is not a significant improvement. We can only constrain $q$ to have upper or lower limits, except for Epoch~7 where $q=4.1^{+1.2}_{-0.8}$; the iron abundance becomes even larger $A_{\rm Fe}=6.3\pm0.7$, and inclination is $42.6^{+1.2}_{-1.7}$ degrees. The most notable change is that the ionization parameter is less well constrained. The inner radius is unconstrained in Epoch~1, slightly larger in Epoch~3 and 4 ($\sim3R_{\rm ISCO}$ compared to $\le2R_{\rm ISCO}$), decreases to $<1.6R_{\rm ISCO}$ in Epoch~6 ($<7.9R_{\rm ISCO}$ in the $q=3$ fit), and matches the $q=3$ fit in the other 3 epochs. 


With a lamppost geometry modeled by \texttt{relxilllpCp}, PG-stat is larger than the $q=3$ fit by 76 with the same number of d.o.f. The iron abundance drops slightly from $\sim4$ to $\sim3$, and also the inclination ($\sim38$ to $\sim32$ degrees). We confirm the same trend in evolution with luminosity the disk component and ionization. The predicted reflection fraction under the lamppost geometry also evolves with luminosity, and is shifted to larger values (e.g., $\sim1.2$ compared to $\sim0.13$ in Epoch~3), which could help relax the requirement for the high coronal luminosity. Contour plots suggest that the lamppost height $h$ is highly degenerate with the inner radius of the disk, so we are not able to determine these parameters independently from our data.


\section{Time lag analysis with \nicer\ data} \label{timing}
\begin{figure*}[htb!]
\centering
\includegraphics[width=1.\linewidth]{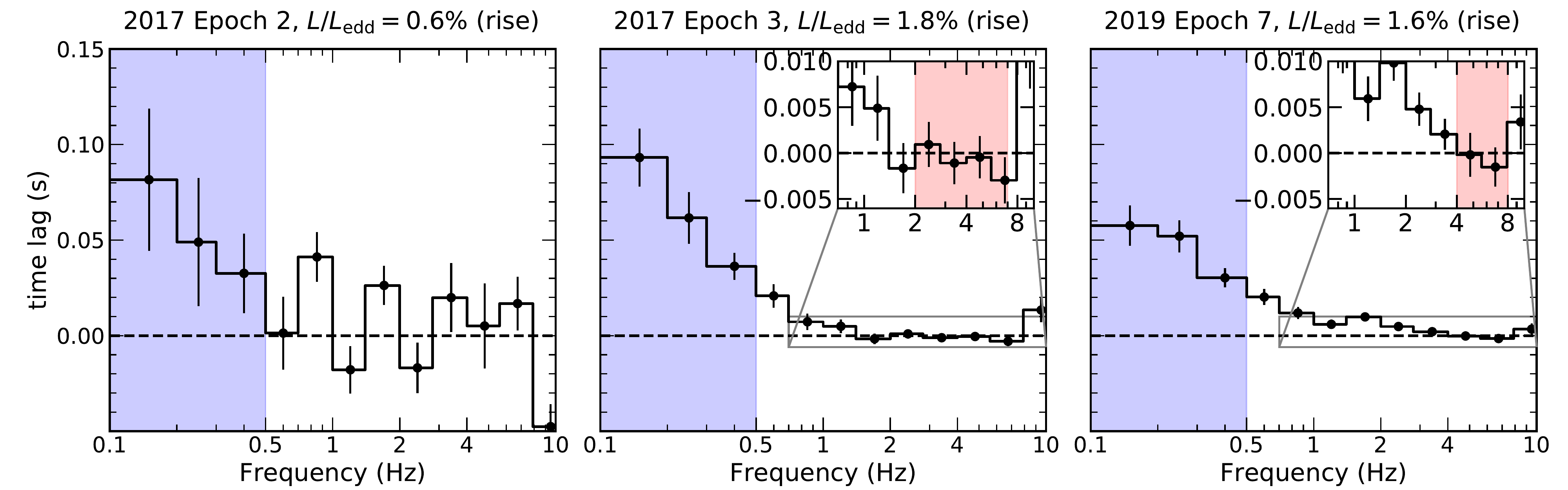}
\caption{The lag-frequency spectra in the 3 epochs with \nicer\ data. Poisson noise dominants above 10~Hz. The shaded frequency range in blue ($0.1-0.5$~Hz) is used to generate the low-frequency lag-energy spectra, resulting in a log-linear dependence of lag on energy in Epoch~3 and 7. The shaded frequency ranges in the inserted zoomed-in lag-frequency spectra in red are where potential candidates of a soft lag from reverberation are present. Those frequency ranges are used to generate the high-frequency lag-energy spectra in Fig.~\ref{fig:lag2}. 
}
\label{fig:lag1}
\end{figure*}

As introduced in Section~\ref{introduction}, \nicer\ is also a great instrument to conduct timing analysis, so we also explore the \nicer\ data from a timing perspective. The epoch indexing is the same as in the spectral analysis, but here we only consider epochs with \nicer\ coverage, namely, Epochs~1, 2, 3, 5 and 7. 

In the $0.3-10$~keV Poisson noise-subtracted power-density spectra, no quasi-periodic oscillations (QPOs) are detected. This is as expected because a systematic QPO search \citep{motta2011low} for \gx339 found no QPO in the low-luminosity hard state. We notice that for Epochs~1 and 5 where the luminosity is lowest, Poisson noise becomes dominant above $\sim0.5$~Hz, meaning that the signal-to-noise ratio is too low to extract any convincing results. In Epochs~2, 3, and 7, on the other hand, Poisson noise can be safely ignored at frequencies below 10, 30, and 20~Hz respectively. 

Following standard Fourier timing techniques \citep{uttley2014x}, we then calculate the cross-spectrum in each epoch, between $0.5-1$~keV and $1-10$~keV energy bands, to obtain the hard-to-soft lag spectra as a function of frequency. Using a standard logarithmic frequency rebinning with a factor of 0.4, the lag-frequency spectra in the Epochs~2, 3, and 7 are shown in Fig.~\ref{fig:lag1}, where we use the convention that a positive lag indicates a hard lag, meaning that the hard photons lag behind the soft photons. In all 3 epochs where Poisson noise can be safely ignored at frequencies below 10~Hz, a low-frequency hard lag is present, with its amplitude decreasing with frequency. This low-frequency hard lag can be fitted with a phenomenological power-law model \citep{nowak1999rossi}, and the indices are, in order of increasing luminosity: $-0.6^{+0.3}_{-0.2}$ (Epoch~2), $-1.5^{+0.1}_{-0.2}$ (Epoch~3), and $-0.9\pm0.1$ (Epoch~7), suggesting that the low-frequency hard lag decreases with frequency faster when the luminosity is higher. The same trend was also found in a previous time lag analysis work on \gx339\ using \xmm\ data \citep{de2015tracing}. At high frequencies in Epoch~3 ($2-7$~Hz) and 7 ($4-8$~Hz), we observe hints of soft lags that could be due to reverberation with millisecond amplitudes. 



The energy-resolved lag spectra are obtained by calculating the cross-spectrum between each energy bin and a reference band (chosen to be $0.5-10$~keV), following standard techniques in \citealt{uttley2014x}. From the low-frequency ($0.1-0.5$~Hz) lag-energy spectra in the 3 epochs where hard lags are confidently found, we see lags with large uncertainties in Epoch~2 limited by the low signal-to-noise; whereas in Epoch~3 and 7, we find a log-linear dependence of lag on energy, but no clear difference between the epochs can be determined, considering the statistical errors. 

\begin{figure}[htb!]
\centering
\includegraphics[width=1.\linewidth]{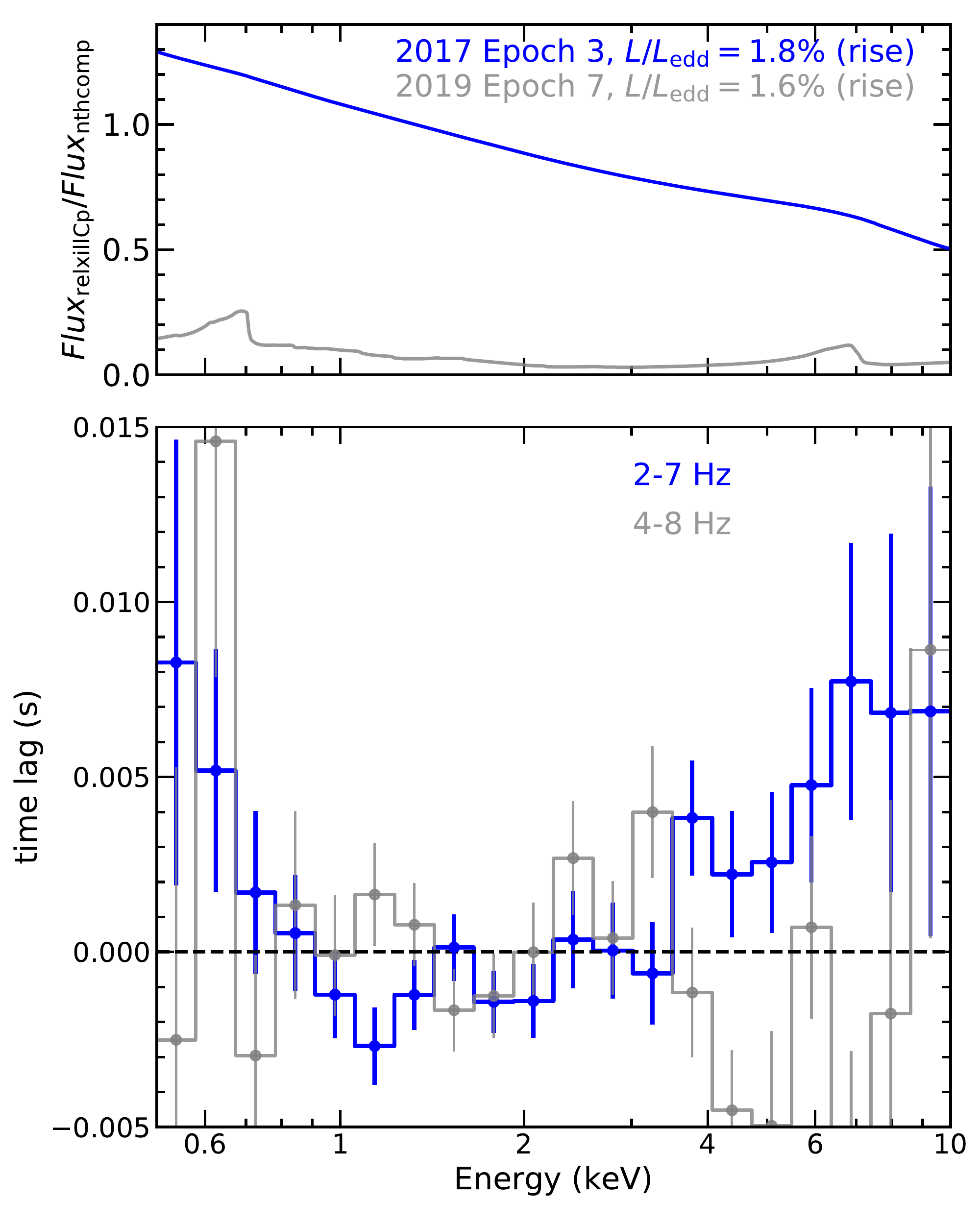}
\caption{(\textit{Upper}) The energy-resolved reflection fraction (reflection/continuum), to illustrate the possibility that the reverberation lag is undetectable in Epoch~7 (\textit{grey}) because the ratio of flux from the reflection component to the continuum is so much smaller than in Epoch~3 (\textit{red}), so that the dilution effect becomes larger. (\textit{Lower}) The high-frequency lag-energy spectra in the frequency range of $2-7$~Hz in Epoch~3, and $4-8$~Hz in Epoch~7. Tentative thermal and iron K reverberation lags are seen in Epoch~3. See Section~\ref{timing} for more details. 
}
\label{fig:lag2}
\end{figure}

At high frequencies where we see the potential soft lags, we can use the lag-energy spectra to explore the reverberation picture. Fig.~\ref{fig:lag2} (\textit{lower}) shows the high-frequency lag-energy spectra in Epoch~3 ($2-7$~Hz) and Epoch~7 ($4-8$~Hz). In Epoch~3, we see tentative hints of a thermal reverberation lag below 1~keV. 

Since the uncertainties are quite large, it remains necessary to test the significance of the reverberation lag. If we fit the lag-energy spectrum with a power-law model with $\Gamma=0$ (fixed), assuming that the hard lag due to propagating fluctuations can be safely ignored at these high frequencies, then the excess at 6.4~keV can be interpreted as an iron K lag. However, considering the large amplitude and uncertainty of the lag, we take the conservative approach and use a power-law model with a free $\Gamma$ as the null hypothesis ($\Gamma=1.0\pm0.4$ and $\chi^2/d.o.f.=19.6/18=1.1$). Under this null hypothesis, the iron K lag is no longer significant ($<1\sigma$ using F-test), i.e., a non-detection. 

For the thermal reverberation lag, it could be modeled with a \texttt{diskbb} if it results from re-thermalizing of the disk, and/or a \texttt{laor} model if from smeared iron~L line. With our data, we can not distinguish these two cases, but in either case, the significance of the thermal reverberation lag is above $2.5\sigma$ using F-test. For consistency with the previous time lag analysis for the 2015 outburst with \xmm\ data \citep{de2017evolution}, we estimate the thermal reverberation lag amplitude to be the maximum intensity of residuals above the extrapolation of the power-law model. The thermal reverberation lag found in this way is $9\pm3$~ms, which is consistent with the result therein. For instance, their highest luminosity is slightly lower than that in Epoch~3 ($1.6\%$ compared to $1.8\%$ in Eddington units), and the thermal reverberation lag is $8\pm3$~ms in that observation.

Epoch~7 reaches a luminosity slightly lower than Epoch~3 ($1.6\%$ versus $1.8\%L_{\rm Edd}$), with a count rate lower by a factor of $\sim1.3$, but its effective exposure used in timing with segment length of 10~s is larger by a factor of $\sim2.3$. For BHBs in the reverberation frequency region, the signal-to-noise of lag measurements scales with count rate and the square root of the effective exposure \citep{uttley2014x}, resulting in a signal-to-noise of lag $\sim1.2$ times larger in Epoch~7. In other words, we might statistically expect a soft X-ray lag in this epoch as well. However, as shown in the lag-energy spectrum of Epoch~7 (Fig.~\ref{fig:lag2}, \textit{lower} panel), there is no reverberation signature. This may be a consequence of the fact that the reflected photons represent a much smaller fraction of the X-ray emission in Epoch~7 than in Epoch~3 (see upper panel of Figure~\ref{fig:lag2} where the flux ratio of the reflection and continuum components are shown). They are different by a factor of $\gtrsim4$ both below 1~keV where thermal reverberation could be seen and at around 6.4~keV where iron K reverberation could be present. A small fraction of reflected photons would lead to a larger dilution of reverberation lag, making it more difficult to observe. Also note that the ratio of flux from the reflection component to the continuum is related to -- but distinct from -- the reflection fraction $R_f$ in Table~\ref{tab:parameters} and Fig.~\ref{fig:par_evol}, which is defined as the ratio of coronal intensity that illuminates the disk and reaches the observer at infinity (see \citealp{dauser2016normalizing}).



\begin{figure*}[htb!]
\centering
\includegraphics[width=1.\linewidth]{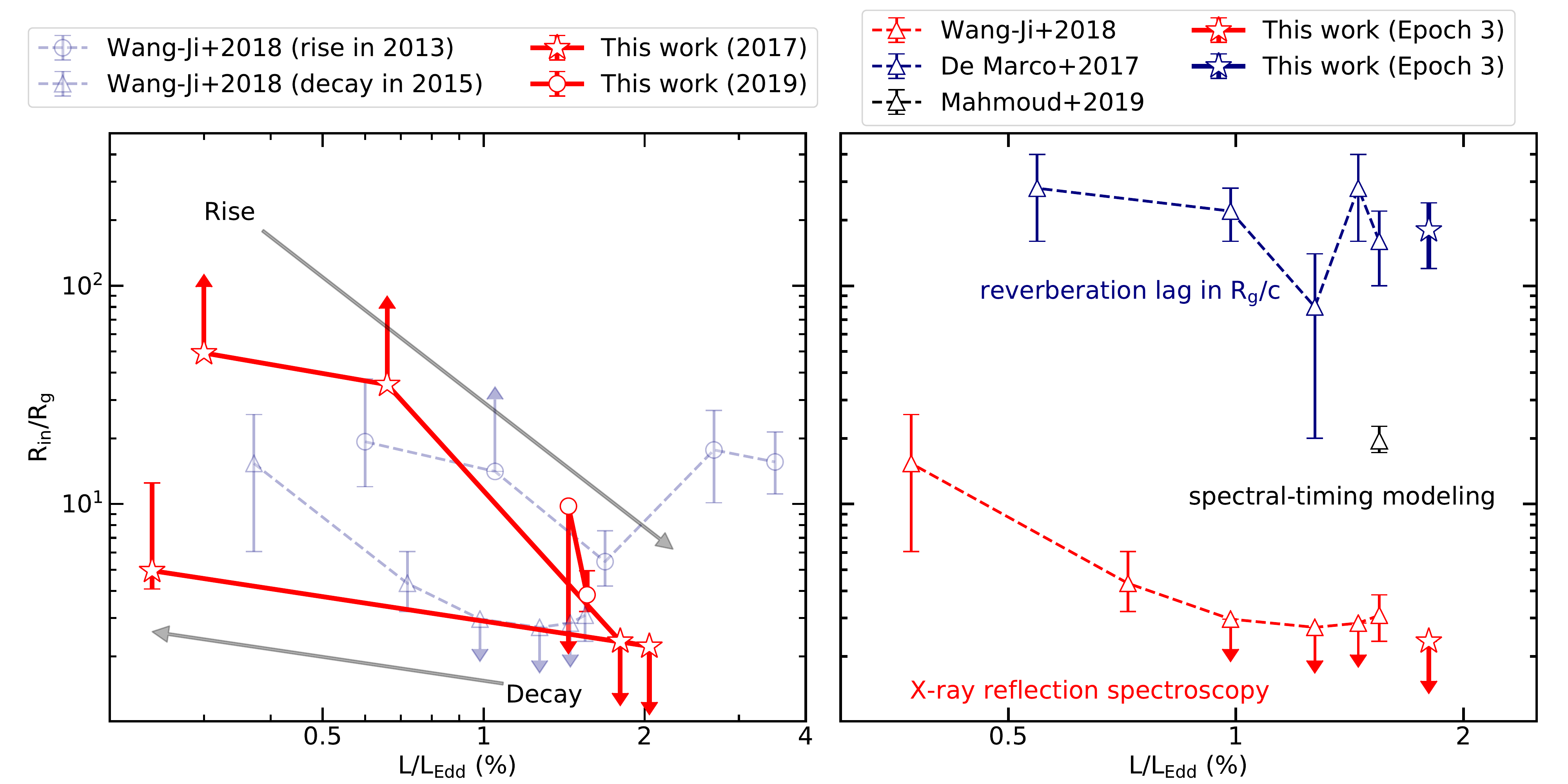}
\caption{(\textit{Left}) Comparison of the inner-disk radius vs. Eddington-scaled luminosity
for \gx339. Our best fit values from reflection spectroscopy are shown in contrast with \citet{wang2018evolution}.  Notice that $R_{\rm in}$ values in Table~\ref{tab:parameters} are in unit of $R_{\rm ISCO}$, so a factor of 1.23 ($a_*=0.998$) is corrected to be in $R_g$ unit. Two arrows show the evolution trend in rise and decay phases, suggesting a hysteresis effect. (\textit{Right}) Comparison of $R_{\rm in}$ obtained from reflection spectroscopy and the thermal reverberation lag amplitude in $R_g/c$ unit, from the analysis of the 2015 outburst data \citep{de2017evolution,wang2018evolution} and this work. The Eddington ratio here assumes the unabsorbed flux in 1-100~keV, $d=8$~kpc, and $M_{\rm BH}=10M_\odot$, i.e., the values in other works have been converted accordingly. }
\label{fig:hysteresis_spectral_timing}
\end{figure*}

\section{Discussion} \label{discussion}

\subsection{Evolution of inner disk radius and possible hysteresis effects}

We found that over the luminosity range of 0.3\% to 2.1\%~$L_{\rm Edd}$, the inner disk radius moves inwards, and the trends of disk component flux, the ionization, and the reflection fraction, are consistent with what are expected. This coincides with an accretion flow increasing its radiative efficiency above $\sim1\%$~$L_{\rm Edd}$, which is both theoretically anticipated (e.g., see \citealp{yuan2014hot}, Section 2.5) and supported by the change of slope in the radio/X-ray correlation \citep{coriat2011radiatively,koljonen2019radio}. However, there are a few surprises from the spectral analysis.

From Figure~5, the photon index $\Gamma$ does not show a clear correlation with the inner edge of the disk $R_{\rm in}$, as would be expected if considering only the amount of cooling corona gets from thermal disk photons. However, $\Gamma$ is governed not only by the coronal temperature, but also the coronal optical depth \citep{lightman1987pair}. As mentioned in Sec.~3.1, our final fit assumes a fixed coronal temperature at 400~keV because previous endeavors found a pegged coronal temperature at that maximal value. This assumption limits us from obtaining estimates of the optical depth. In \citet{javier_gx339}, where the coronal temperature and optical depth are both constrained, $\Gamma$ does not show a simple correlation with $R_{\rm in}$. The situation becomes even more complicated considering the unknown heating mechanism of the corona, i.e., how the accretion power gets dissipated into heating the corona to such high temperatures. We regard this as an important point, and one that needs to be reconciled in most physical pictures for the evolution of the inner edge of the accretion disk in the hard outburst. 


We also notice that the change of the normalization of \texttt{xillverCp} ($N_{\rm x}$) is faster than the change of total X-ray flux. For example, when comparing Epoch~4 and 5, the X-ray luminosity decreases by a factor of 7, while $N_{\rm x}$ decreases by a factor larger than 15. This could be due to an intrinsic change of coronal geometry including both its size and location, irradiating a different solid angle of the disk at large radii. By generating contours between $N_{\rm x}$ and $R_{\rm in}$ using \texttt{STEPPAR} command in \texttt{XSPEC}, we confirm that those two parameters are not correlated, so that our results are not affected. 

With our new measurements of $R_{\rm in}$, we can extend the reported results on the evolution of $R_{in}$ with luminosity in the hard state (e.g., \citealt{javier_gx339, wang2018evolution, garcia20192017}). The new results derived from \nustar\ and \nicer, both free from pile-up, agrees with the picture that the inner radius moves inwards as the luminosity increases in the hard state: in our 4 epochs with $L>1\%L_{\rm Edd}$, $R_{\rm in}\le10R_g$, indicating that any truncation from the ISCO must be quite minor. At lower luminosities, as we mentioned in Section~\ref{evolution}, the disk appears closer in during the decay in comparison to the rise, suggesting a hysteresis effect. This idea is also supported by \citet{wang2018evolution}, where we found that $R_{\rm in}$ was smaller in 2015 (decay) than in 2013 (rise) in the shared luminosity range (see the left panel in Fig.~\ref{fig:hysteresis_spectral_timing}). 

We stress that even though the reflection spectra, especially the iron line shapes, are notably different in Epochs~1 and 5 during the rise and decay (see Fig.~\ref{fig:ratio_po}), hysteresis is only marginally significant ($\Delta$PG-stat = 8 for 1 extra d.o.f.). Also, in the fit under the lamppost geometry, no hysteresis effect could be seen, but the PG-stat is worse by 76 with the same number of d.o.f. However, the simultaneous change in ionization and reflection fraction, all suggest a hysteresis picture. If the hysteresis effect is real, possible physical interpretations include: (1) different accretion modes at different mass accretion rates affected by the amount of Compton cooling or heating on the accretion disk \citep{meyer2005hysteresis}; (2) the interplay of magnetic fields and the accretion flow \citep{petrucci2008role, begelman2014mechanism, cao2016accretion}; (3) thermal limit cycles \citep{latter2012hysteresis}. More spectral-timing analysis for data with similar luminosity during the rise and decay is needed to further explore the hysteresis effect. 


The paradigm where there is only a small level of disk truncation above $\sim1\%L_{\rm Edd}$ is challenged by observational evidence from both spectral and timing sides. From the spectral side, the focus has shifted from pile-up effects to model dependence: different models for reflection and the underlying continuum can lead to different inferred levels of disk truncation \citep{dzielak2019comparison}. Using a high density reflection model on the same dataset as in \citet{wang2018evolution}, \citet{jiang2019high} found higher upper limits on $R_{\rm in}$ for several epochs, e.g., $R_{\rm in}<8R_{\rm ISCO}$ rather than $R_{\rm in}=1.8^{+3.0}_{-0.6}R_{\rm ISCO}$ in an epoch with $L=1.4\%L_{\rm Edd}$. \citet{mahmoud2019reverberation} uses a double Comptonization model to account for temperature gradient in the inner-hot-flow-like corona. We have tested both the high density reflection model and the double Comptonization model for Epoch~3 and found the fit statistics were not as good as our final model. More details can be found in Appendix.




\subsection{Understanding time lags and the need for spectral timing modelling}

From the timing side, one aspect of debate comes from converting the reverberation lag amplitude to a light travel time delay, i.e., between the corona and the reflecting inner disk. We can use our constraints on the amplitudes of thermal reverberation lags as an example to illustrate this (see Fig.~\ref{fig:hysteresis_spectral_timing}). In the most naive picture, the lag corresponds to a distance of $d=\tau c/(1+cos[\theta-i])$ where $\tau$ is the measured lag, $\theta$ is the angle defining the location of the corona relative to the reflector ($\theta=0$ represents a corona in the lamppost geometry and $\theta=\pi/2$ represents a central corona in a truncated disk geomtry). Assuming a black hole mass of $10M_\odot$ and an inclination angle $38^\circ$ as we found with spectral fitting, a reverberation lag amplitude of 1~ms corresponds to a light travel time of $\sim12R_g/c$ regardless of the exact value of $\theta$. Then, our average thermal reverberation lag, $9\pm3$~ms, would correspond to a corona to inner disk distance of $108\pm36$~$R_g/c$. If considering dilution effects, the intrinsic lag would even be larger. While for the same epoch (Epoch~3), $R_{\rm in}$ determined from energy spectral fitting is $<2.4R_g$, leading to a large discrepancy. 



The very large inferred truncation radius from reverberation lag amplitude is extreme and suggests that a direct conversion from lag amplitude to light travel distance between corona and inner disk is not prudent. The primary reason is that even though the response function peaks at the disk truncation radius, which should be the same as inferred from the spectral fitting, it has a wing towards longer lags (i.e., the other parts of the disk also contribute to the measured ``average" lag). In addition, there are secondary caveats. As we discussed in Section~\ref{timing}, we cannot tell if any energy-dependence in the hard lag is still present in the ``high" frequency range we adopt ($2-7$~Hz). It is therefore possible that limited by the signal-to-noise ratio, we are not probing frequencies high enough so that reverberation lag becomes dominant. It is also worthwhile to note that in the recent work with \nicer\ data on a bright BHB MAXI~J1820+070 \citep{kara2019corona}, the averaged observed iron K reverberation lag is $0.47\pm0.08$~ms, which is one order of magnitude smaller than ours, and corresponds to $14\pm3R_g/c$. This reverberation lag amplitude is also the smallest measured so far in any BHB. However, the count rate reached by that source is $\sim20,000$ counts/sec, more than 100 times larger than by \gx339 in these faint outbursts.




One promising approach to settle the discrepancy is to conduct self-consistent spectral-timing analysis taking into account a proper transfer function. We notice that the two available spectral-timing models use a different approach to deal with the hard lag whose nature we do not understand yet. \texttt{PROPFLUC} \citep{mahmoud2018modelling,mahmoud2018physical,mahmoud2019reverberation} assumes that the inner disk evaporates into a geometrically thick and hot flow which is radially stratified into a two-temperature continuum. This is one specific geometry of the disk-corona, which makes \texttt{PROPFLUC} a bottom-up approach. On the other hand, the phenomenological hard lag treatment with the pivoting power-law in the newly public model \texttt{reltrans} \citep{ingram2019public, mastroserio2019x} is a top-down approach, using which different theoretical predictions could be tested. 

With \texttt{reltrans}, we could fit jointly the time-averaged spectrum and the real and imaginary parts of the energy-dependent cross-spectrum for a range of Fourier frequencies. The first X-ray reverberation mass measurement of a stellar-mass black hole is obtained with \texttt{reltrans} for Cygnux~X-1 \citep{mastroserio2019x}. The fitted disk truncation radius is $<10R_g$ at $\sim$1.6\%~$L_{\rm Edd}$. 

With the \texttt{PROPFLUC} model, \citet{mahmoud2019reverberation} found that a truncation of $\sim20R_g$ could explain the lag-energy spectra in three frequency bands and the power spectral densities (PSDs) in three energy bands. The observation used is the highest luminosity observation in the 2015 outburst ($1.6\%L_{\rm Edd}$). The thermal reverberation lag amplitude is $8\pm3$~ms ($96\pm36$~$R_g/c$ in light travel time, assuming $M_{\rm BH}=10M_\odot$ and inclination of 38 degrees). For the same observation, but using \swift\ instead of \xmm\ for soft energy coverage, reflection spectral fitting yields $R_{in}<4R_g$ with 90\% confidence \citep{wang2018evolution}. Even though the result from spectral-timing still gives a large truncation of 20$R_g$, it is much smaller than the reverberation lag in $R_g/c$ unit. It is much closer to the results from reflection spectroscopy, but still shows discrepancy. \\



\section{Summary}

We have analyzed 7 epochs of \gx339\ in the hard state seen by
\nicer\ and \nustar\,, 5 taken in the 2017 outburst and the other 2
during the rise of the 2019 outburst, with both outbursts being hard-only. The data cover a luminosity range of 0.3\% to 2.1\% $L_{\rm Edd}$. Our major findings are as follows: 

\begin{enumerate}[label=(\roman*), wide, labelwidth=!, labelindent=0pt]
\item During the rise in 2017, the inner disk moves towards the ISCO radius, from $>49R_g$ to $<2R_g$. Other physical quantities show an evolution consistent with the $R_{\rm in}$ evolution, including a larger unscattered flux contribution from the thermal disk component (along with a higher disk temperature up to $\sim200$~eV), an increasing ionization parameter, and an increasing reflection fraction (see Fig.~5). 

\item Epoch~5 has a similar luminosity as Epoch~1 but occurs during the decay; this observation shows a mild truncation at $\sim5R_g$, while the reflection fraction is $\sim$4~times larger, indicating a hysteresis effect during the rise and decay (see Fig.~8, left). The hysteresis effect has several physical interpretations, and is potentially important for us to understand the accretion process and the mechanisms governing state transitions. 

\item In our 4 epochs with $L>1\%L_{\rm Edd}$, we find that the disk is at most slightly truncated ($R_{\rm in}\le10R_g$). 

\item We observe a tentative thermal reverberation lag, with an amplitude of $9\pm3$~ms, consistent with previous findings using \xmm\ data. We also discussed possible reasons for the large discrepancy between the disk truncation level determined from spectral modeling and the large reverberation lag amplitude assuming it comes from light travel time delay between the corona and the inner disk. Possible reasons include contributions from reverberation from outer parts of the disk, the hard lags and the soft-excess interpretation. Future spectral-timing modeling is needed to settle the discrepancy. 

\end{enumerate}

\bigskip
JW, EK and JAG acknowledges support from NASA grant 80NSSC17K0515, and thank the International Space Science Institute (ISSI) and participants of the ISSI Workshop ``Sombreros and lampposts: The Geometry of Accretion onto Black Holes" for fruitful discussions. JAG thanks support from the Alexander von Humboldt Foundation. EMC gratefully acknowledges support from the National Science Foundation, CAREER award number AST-1351222. RML acknowledges the support of NASA through Hubble Fellowship Program grant HST-HF2-51440.001.

\appendix

As discussed in Section~5.1, the choices of reflection model and underlying continnum shape can possibly change the result of disk truncation radius. Therefore, we tested the double Comptonization \citep{mahmoud2019reverberation} and the high density reflection model \citep{jiang2019high} for Epoch~3, which has the highest quality spectra. Our best-fit model is M1: \texttt{Tbabs*crabcorr*(diskbb+relxillCp+xillverCp+gaussian)*edge*edge}, the double Comptonization model is M2: \texttt{Tbabs*crabcorr*(diskbb+relxillCp+relxillCp+gaussian)*edge*edge}, and the high density reflection model is \texttt{Tbabs*crabcorr*(relxillD+xillverD+gaussian)*edge*edge}. 

The treatment in M1 is the same as in our manuscript. For M2, $kT_{\rm e, hard}$=100~keV, $kT_{\rm e, soft}$=35~keV as assumed in \citet{mahmoud2019reverberation}; in the second ``soft" reflection component, only the normalization, photon index, and reflection fraction are untied from the ``hard" reflection component. For M3, the parameters in \texttt{xillverD} are tied to \texttt{relxillD}, except for that ionization is assumed to be $log\xi=0$, and the normalization is free to vary. 

The resulting PG-Stat./d.o.f. is $1690.0/1556=1.08$ for M1, $1721.9/1554=1.11$ for M2, and $1759.2/1557=1.13$ for M3. We notice that with M0: \texttt{Tbabs*crabcorr*(diskbb+relxillCp+gaussian)*edge*edge}, PG-Stat./d.o.f. is $1777.7/1557=1.12$, so M2 reduced PG-Stat by 55.8 with 3 more d.o.f., which is a great improvement, but is still not as good as M1, which reduces PG-Stat by 87.7 with only 1 more d.o.f. We also check that if an extra distant reflection model component \texttt{xillverCp} is added to M2, the statistics is not improved, its normalization is $<3.4\times10^{-5}$, and it does not affect other parameters. 

The best-fit parameters are shown in Table~\ref{tab:M1M2}, the model components and data-to-model ratios in Figure~\ref{fig:M1M2}. For M1, the individual fitting results are consistent with those shown in Table~\ref{tab:parameters} for the simultaneous fitting in the manuscript, except for that the emissivity index $q>4.9$ here while we fix it at 3 in the manuscript. However, fixing $q$ at 3 here would only increase the PG-Stat. by 3.3, which means the spectrum is insensitive to properly constrain $q$, and the value of $q$ would not affect the fit. As for M2, the disk truncation level ($R_{\rm in}=2.95^{+0.10}_{-0.09}$~$R_{\rm ISCO}$) is similar with M1, instead of that it is more largely truncated as suggested in \citet{mahmoud2019reverberation}. Also, the double Comptonization model does not return a closer-to-solar iron abundance. In fact, iron abundance is increased in M2 ($\sim8.8$) compared to M1 ($\sim4.5$). 

Regarding M3, the disk component is no longer required because high density reflection could add a quasi-blackbody emission in the soft band below 2~keV as suggested in \citet{jiang2019high}. The inclination is $\sim70$ degree, which is much higher than in M1 and M2, and the spectrum is much harder with $\Gamma\sim1.36$. The iron abundance is pegged at maximal value of 10, and $R_{\rm in}<1.01$~$R_{\rm ISCO}$; both trends are inconsistent with \citet{jiang2019high}. However, we notice that in M3 here, we use the high density reflection model \texttt{relxillD}, which means that we are limited by the higher limit of density of $\log (N/{\rm cm}^{-3}) =19$, and the fitted density is pegged at this higher limit ($\log (N/{\rm cm}^{-3}) >18.7$); while \citet{jiang2019high} found $\log (N/{\rm cm}^{-3}) =20.6$ with the reflection model \texttt{reflionx}. Therefore, the changes of iron abundance and disk inner radius when high density is considered still needs investigation with more atomic data with high electron densities. 

\setcounter{table}{0}
\renewcommand{\thetable}{A\arabic{table}}
\begin{table}[htb!]
\begin{center}
\caption{Best fit parameters for our best-fit model M1, double Comptonization model M2 and high density reflection model M3 in Epoch~3 ($1.8\%L_{\rm Edd}$).  \label{tab:M1M2}}
\begin{tabular}{ccccc}\hline \hline
Model component & Parameter & M1 & M2 & M3\\
\hline
\textsc{relxillCp}&$a_*$& \multicolumn{3}{c}{$0.988$ (f)}  \\
\textsc{crabcorr}&$\Delta \Gamma_{\rm NICER}$& \multicolumn{3}{c}{$0$ (f)} \\
\hline
\textsc{Tbabs}&$N_H$ ($10^{21}$cm$^{-2}$)&  $6.32\pm0.17$ & $6.72^{+0.08}_{-0.01}$& $5.55^{+0.08}_{-0.07}$\\
\textsc{diskbb}&$T_{\rm in}$ (eV)& $196^{+13}_{-10}$ & $204^{+20}_{-10}$ & -\\
\textsc{diskbb}& Norm ($10^3$)& $5.7^{+2.9}_{-2.1}$ & $3.2^{+0.2}_{-0.7}$& - \\
\textsc{relxillCp}/\textsc{relxillD}&$q$& $>4.9$ &  $6.0^{+0.9}_{-0.7}$&  $8.0^{+1.5}_{-1.7}$ \\
\textsc{relxillCp}/\textsc{relxillD}&$i$ (degrees)&  $48.7^{+10.3}_{-7.7}$ &  $44.1^{+0.6}_{-0.7}$&  $69.7^{+3.3}_{-6.0}$\\
\textsc{relxillCp}/\textsc{relxillD}&$R_{\rm in}$ ($R_{\rm ISCO}$)& $2.0^{+0.8}_{-0.5}$ & $2.95^{+0.10}_{-0.09}$  & $<1.01$ \\
\textsc{relxillCp}/\textsc{relxillD}&$\Gamma$& $1.51^{+0.02}_{-0.07}$ & $1.5710^{+0.0004}_{-0.0126}$ & $1.364^{+0.024}_{-0.011}$ \\
\textsc{relxillCp}/\textsc{relxillD}&$\log \xi$ (erg$\cdot$cm$\cdot$s$^{-1}$) & $3.80^{+0.10}_{-0.37}$ & $<2.01$  & $3.49\pm0.02$ \\
\textsc{relxillCp}/\textsc{relxillD}&$A_{\rm Fe}$&  $4.5^{+2.0}_{-0.50}$ &  $8.8\pm0.5$&  $>9.5$\\
\textsc{relxillCp}/\textsc{relxillD}&$\log N$ (cm$^{-3}$) & 15 (f)  &  15 (f)&  $>18.7$\\
\textsc{relxillCp}/\textsc{relxillD}&$R_{\rm f}$& $0.12^{+0.08}_{-0.05}$ & $0.127\pm0.003$ & $0.085^{+0.006}_{-0.002}$\\
\textsc{relxillCp}/\textsc{relxillD}&$kT_{\rm e}$ (keV)&$400$ (f) &$100$ (f)& -\\
\textsc{relxillCp}/\textsc{relxillD}&Norm ($10^{-3}$)& $5.4\pm0.6$ & $5.491^{+0.008}_{-0.038}$& $6.2\pm0.3$ \\
\hline
\textsc{xillverCp}/\textsc{xillverD}&Norm ($10^{-4}$)& $6.9^{+2.6}_{-2.3}$ & -& $5.5^{+0.9}_{-1.0}$\\
\textsc{relxillCp$_{\rm soft}$}&$\Gamma_{\rm soft}$& - & $3.20^{+0.04}_{-0.01}$ & -\\
\textsc{relxillCp$_{\rm soft}$}&$R_{\rm f, soft}$& - & (u) & -\\
\textsc{relxillCp$_{\rm soft}$}&$kT_{\rm e, soft}$ (keV)&- &$35$ (f)& -\\
\hline
\textsc{Gaussian}&$E$ (keV)&  $1.85\pm0.04$&  $1.88\pm0.04$ &  $1.91\pm0.04$\\
\textsc{Gaussian}& $\sigma$ (keV)& $>0.089$ & $0.19^{+0.03}_{-0.02}$ & $>0.096$\\
\textsc{Gaussian}& Norm ($10^{-4}$) & $3.6\pm1.0$ & $7.8^{+1.0}_{-0.9}$& $4.7\pm0.9$ \\
\textsc{$\rm edge_1$} & $E$ (keV)& $2.29^{+0.05}_{-0.06}$ & $2.3\pm0.03$& $2.28^{+0.03}_{-0.04}$\\
\textsc{$\rm edge_1$}& $\tau_{\rm max}$ & $-0.023\pm0.011$ & $-0.050\pm0.006$ & $-0.062\pm0.010$ \\
\textsc{$\rm edge_2$}&$E$ (keV)&  $0.532^{+0.007}_{-0.008}$ &  $0.534^{+0.003}_{-0.004}$&  $0.533\pm0.008$\\
\textsc{$\rm edge_2$}& $\tau_{\rm max}$ & $0.42^{+0.07}_{-0.06}$ & $0.415^{+0.046}_{-0.013}$& $0.36^{+0.07}_{-0.06}$\\
\hline
\textsc{crabcorr}&$\Delta \Gamma_{\rm FPMA}$& $-0.018\pm0.015$ & $0.022\pm0.012$ & $0.0186^{+0.014}_{-0.017}$ \\
\textsc{crabcorr}&$N_{\rm FPMA}$& $1.189^{+0.016}_{-0.028}$ & $1.271^{+0.022}_{-0.003}$& $1.26\pm0.03$\\
\textsc{crabcorr}&$\Delta \Gamma_{\rm FPMB}$& $-0.044\pm0.015$ & $-0.0043\pm0.0012$& $-0.008^{+0.012}_{-0.017}$\\
\textsc{crabcorr}&$N_{\rm FPMB}$& $1.15\pm0.03$ & $1.233\pm0.003$& $1.225\pm0.003$\\
\hline
&PG-Stat./d.o.f.& $1690.0/1556=1.08$ & $1721.9/1554=1.11$& $1759.2/1557=1.13$\\
\hline
\end{tabular}
\\
\raggedright{\textbf{Notes.} \\
Errors are at 90\% confidence level and statistical only. }
\end{center}
\end{table}

\setcounter{figure}{0}
\renewcommand{\thefigure}{A\arabic{figure}}
\begin{figure}[htb!]
\centering
\includegraphics[width=1.\linewidth]{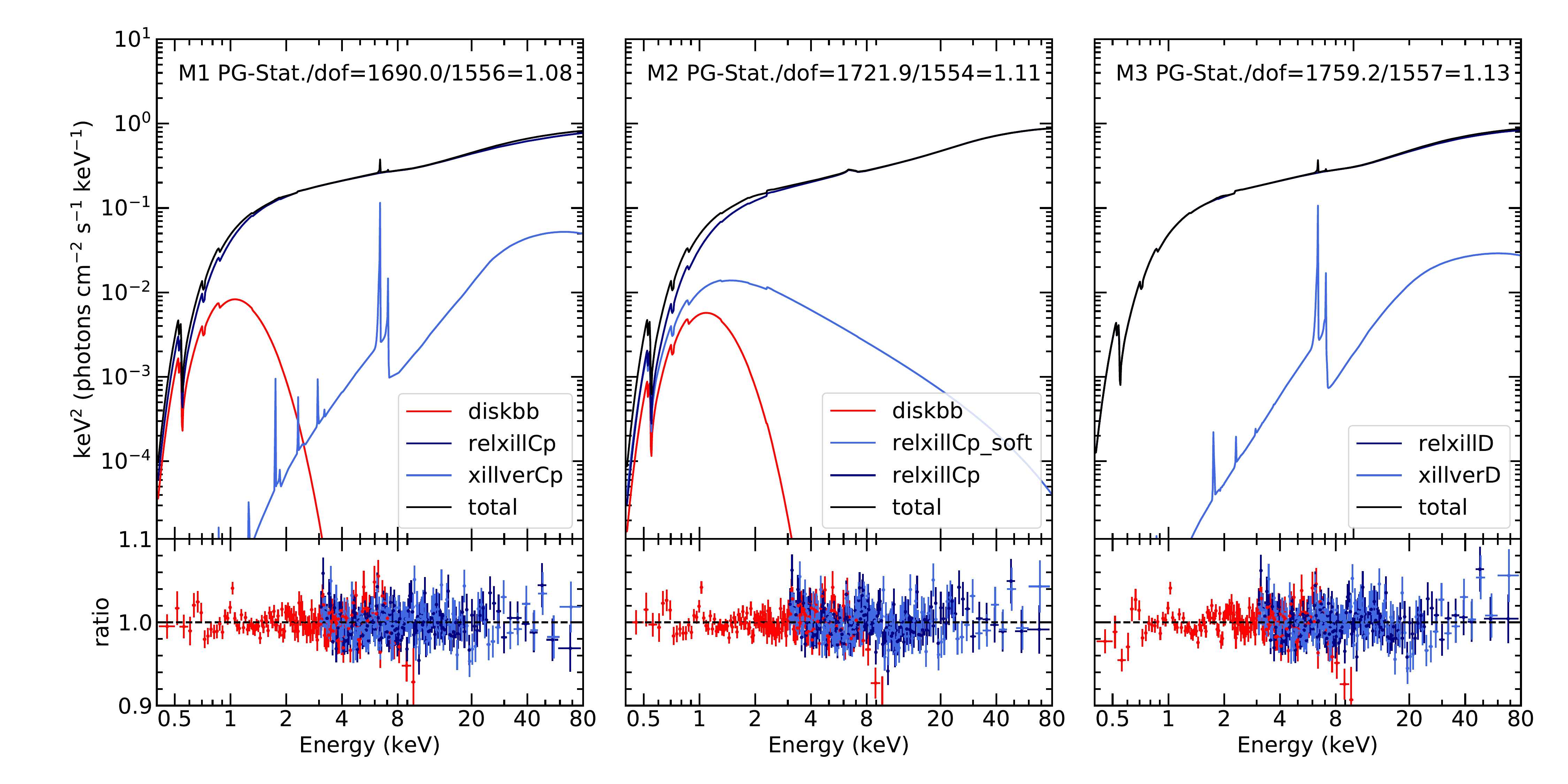}
\caption{Model components (\textit{upper}) and data-to-model ratio (\textit{lower}) for the fit with the our final model M1, the double Comptonization model M2, and the high density model M3. In the ratio plot, red: \nicer\ , navy: \nustar\ /FPMA, blue: \nustar\ /FPMB. \label{fig:M1M2}
}
\end{figure}

\bibliographystyle{apj}
\bibliography{draft-v8}
\end{document}